\documentclass[twocolumn,showpacs,superscriptaddress,aps,pre]{revtex4-1}

\usepackage{amssymb,amsmath}
\usepackage[dvips]{graphicx}

\begin{document}

\title{Contact process on generalized Fibonacci chains: infinite-modulation criticality and
double-log periodic oscillations}

\author{Hatem Barghathi}
\affiliation{Department of Physics, Missouri University of Science and Technology,
Rolla, MO 65409, USA}

\author{David Nozadze}
\affiliation{Department of Physics, Missouri University of Science and Technology,
Rolla, MO 65409, USA}
\affiliation{Department of Physics, The Ohio State University, Columbus, OH 43210, USA}

\author{Thomas Vojta}
\affiliation{Department of Physics, Missouri University of Science and Technology,
Rolla, MO 65409, USA}

\begin{abstract}
We study the nonequilibrium phase transition of the contact process with aperiodic
transition rates using a real-space renormalization group as well as Monte-Carlo
simulations. The transition rates are modulated according to the generalized Fibonacci
sequences defined by the inflation rules A $\to$ AB$^k$ and B $\to$ A. For $k=1$ and 2,
the aperiodic fluctuations are irrelevant, and the nonequilibrium transition is in the
clean directed percolation universality class. For $k\ge 3$, the aperiodic fluctuations
are relevant. We develop a complete theory of the resulting unconventional ``infinite-modulation''
critical point which is characterized by activated dynamical scaling. Moreover, observables
such as the survival probability and the size of the active cloud display pronounced
double-log periodic oscillations in time which reflect the discrete scale invariance of the
aperiodic chains. We illustrate our theory by extensive numerical results, and we discuss
relations to phase transitions in other quasiperiodic systems.
\end{abstract}

\date{\today}

\maketitle


\section{Introduction}
\label{sec:Intro}

Many-particle systems far from equilibrium can display abrupt transitions between
different nonequilibrium steady states that share many characteristics with equilibrium
phase transitions. Examples of such nonequilibrium
phase transitions occur in turbulence, catalytic reactions, interface growth, and
in the dynamics of epidemics and other biological populations
\cite{ZhdanovKasemo94,SchmittmannZia95,MarroDickman99,Hinrichsen00,Odor04,Luebeck04,TauberHowardVollmayrLee05,
HenkelHinrichsenLuebeck_book08}.

Absorbing-state transitions constitute a particularly well-studied subclass of nonequilibrium phase
transitions. They separate active, fluctuating steady states from absorbing states
which are completely inactive and do not display any fluctuations. Generically,
absorbing-state transitions are in the directed percolation (DP) universality class
\cite{GrassbergerdelaTorre79}, provided they feature a scalar order parameter and
short-range interactions but no extra symmetries or conservation laws
\cite{Janssen81,Grassberger82}. The contact process \cite{HarrisTE74} is a
prototypical model in the DP universality class. Experimental examples of absorbing state transitions
were found in turbulent liquid crystals \cite{TKCS07}, periodically driven
suspensions \cite{CCGP08,FFGP11}, and in systems of superconducting vortices \cite{OkumaTsugawaMotohashi11}.

Many realistic experimental systems contain various types of spatial inhomogeneities.
For this reason, the effects of such inhomogeneities on absorbing state transitions have
attracted considerable attention. Random disorder was shown to destabilize the
clean DP critical point \cite{Janssen97} because its correlation
length critical exponent $\nu_\perp$ violates the Harris criterion \cite{Harris74}
$d\nu_\perp>2$ in space dimensions $d=1,2$ and 3. Early numerical simulations
of the disordered contact process
\cite{BramsonDurrettSchonmann91,MoreiraDickman96,*DickmanMoreira98,WACH98,CafieroGabrielliMunoz98}
showed unusually slow dynamics but the ultimate fate of the transition was only
resolved by means of a strong-disorder renormalization group analysis
\cite{HooyberghsIgloiVanderzande03,*HooyberghsIgloiVanderzande04}
of the one-dimensional disordered contact
process. It yielded an
exotic infinite-randomness critical point accompanied by power-law Griffiths
singularities \cite{Noest86,*Noest88}. The renormalization group predictions were
confirmed by Monte-Carlo simulations \cite{VojtaDickison05}, and analogous behavior was
also found in two and three dimensions \cite{OliveiraFerreira08,VojtaFarquharMast09,*Vojta12}
as well as in diluted systems at the lattice percolation threshold \cite{VojtaLee06,*LeeVojta09}.

Spatial inhomogeneities can arise not just from random disorder but also from
deterministic but aperiodic (quasiperiodic) modulations of the transition rates
defining the nonequilibrium process. The stability of a clean critical point against
such aperiodic fluctuations can be tested by means of a generalization of the Harris
criterion, the Harris-Luck criterion \cite{Luck93a}, which relates the clean correlation
length exponent $\nu_\perp$ and the wandering exponent $\omega$ of the aperiodic
structure.

In this paper, we use a real-space renormalization group as well as Monte-Carlo simulations
to study the one-dimensional contact process with aperiodic transition rates
modulated according to the generalized Fibonacci sequences defined by the inflation
rules A $\to$ AB$^k$ and B $\to$ A.
For $k=1$ and 2, the aperiodic fluctuations are irrelevant according to the Harris-Luck
criterion. Correspondingly, we find the nonequilibrium transition to be in the
clean directed percolation universality class. For $k\ge 3$, the aperiodic fluctuations
are relevant. We develop a complete theory of the resulting ``infinite-modulation'' critical
point. It is characterized by a diverging strength of the inhomogeneities and
features activated dynamical scaling similar to the disordered contact
process. Moreover, observables display double-log periodic oscillations in time which reflect the discrete
scale invariance of the aperiodic chains. We also confirm and illustrate the renormalization
group predictions by extensive numerical simulations.

The paper is organized as follows. In Sec.\ \ref{sec:CP}, we introduce the contact process
and the generalized Fibonacci chains. We also discuss the Harris-Luck criterion.
The renormalization group theory is developed in Sec.\ \ref{sec:RG}. Section
\ref{sec:MC} is devoted to the Monte-Carlo simulations. We conclude in Sec.\ \ref{sec:Conclusions}.

\section{Contact process on aperiodic chains}
\label{sec:CP}
\subsection{Generalized Fibonacci chains}
\label{subsec:Fibonacci}

We consider a family of aperiodic two-letter sequences generated by the inflation rules
\begin{equation}
\begin{array}{lll}
  \textrm{A} & \to & \textrm{AB}^k\\
  \textrm{B} & \to & \textrm{A}
\end{array}
\label{eq:inflation}
\end{equation}
where $k$ is a positive integer and  B$^k$ stands for a sequence of $k$ letters B.
The case $k=1$ corresponds to the famous Fibonacci sequence. For $k=2$, the fourth-generation
sequence (starting from a single letter A) reads \mbox{ABBAAABBABB}.
In general, the sequences created by eq.\ (\ref{eq:inflation})  contain groups of $k$ letters B
separated by either single letters A or groups of $k+1$ letters A.
Many properties of
these sequences can be obtained from the substitution matrix
\begin{equation}
\mathbf{M}_k =  \left(\begin{array}{cc}
                        1 & 1 \\
                        k & 0 \\
                      \end{array}
                    \right)
\label{eq:subsmatrix}
\end{equation}
which describes how the numbers $N_A$ and $N_B$ of letters A and B evolve under the inflation
(see, e.g., Ref.\ \cite{Moody97} and references therein). Its eigenvalues read
\begin{equation}
\zeta_\pm = \frac 1 2 \left (1 \pm \sqrt{1+4k} \right)~.
\end{equation}
The larger eigenvalue $\zeta_+$ controls how the total length $N(i) = N_A(i) + N_B(i)$
increases with the inflation step $i$. In the limit of large $i$, one obtains
$N_i \sim \zeta_+^i$. The smaller eigenvalue $\zeta_-$ governs the fluctuations of
the numbers $N_A$ and $N_B$. Specifically, $\Delta N_A (i) =|N_A(i)-x_A N(i)| \sim |\zeta_-|^i$
for large $i$. Here $x_A =\lim_{i\to\infty} N_A(i)/N(i)$ is the fraction of letters A in the
infinite chain. The same relation also holds for $N_B$.
The wandering exponent $\omega$ relates the fluctuations to the length of the chain,
$\Delta N_A(i) \sim N(i)^\omega$. This yields the equation
\begin{equation}
\omega = {\ln|\zeta_-|} / {\ln \zeta_+}~.
\label{eq:wandering}
\end{equation}
For the generalized Fibonacci chains defined in (\ref{eq:inflation}),
the specific values are $\omega_1 = -1$, $\omega_2 = 0$ and $\omega_3 \approx 0.3171$
for $k=1,2$ and 3. Upon further increasing $k$, $\omega$ increases monotonically and
reaches 1 for $k\to \infty$.

\subsection{Contact process}
\label{subsec:CP}

The (clean) contact process \cite{HarrisTE74} is one of the simplest systems undergoing an
absorbing state transition. It can be understood as model for the spreading of an epidemic.
Each lattice site can be in one of two states, active
(infected) or inactive (healthy). Over time, active sites can infect their neighbors or they
can heal spontaneously.  More precisely, the time
evolution is a continuous-time Markov process during which infected sites
heal at a rate $\mu$ while healthy sites become infected by their neighbors
at a rate $\lambda n /(2d)$.
Here, $n$ is the number of sick nearest neighbors of the given site.

The long-time behavior of the system is controlled by the ratio of the
infection rate $\lambda$ and the healing rate $\mu$.
For $\lambda \ll \mu$, healing dominates over infection, and the epidemic eventually dies
out completely. Thus, the model ends up in the absorbing state without
any infected sites. This is the inactive phase. In contrast,
the density of infected sites remains nonzero in the long-time limit if the
infection rate $\lambda$ is sufficiently large, i.e., the model is in the active phase.
The nonequilibrium transition separating these two phases belongs to the DP universality
class.

\begin{figure}
\includegraphics[width=8.5cm]{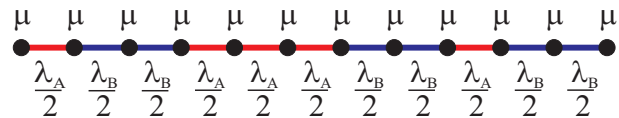}
\caption{(Color online) Sequence of transition rates for the contact process on a generalized Fibonacci chain,
         showing the 4th generation of the $k=2$ chain.}
\label{fig:chain}
\end{figure}

Spatial inhomogeneity can be introduced into the contact process by making the infection
and/or healing rates dependent on the lattice site. We are interested in aperiodic
(quasiperiodic) inhomogeneities. Specifically, we consider a chain of sites that have equal healing
rates $\mu$ but two different the infection rates $\lambda_A$ and $\lambda_B$
\footnote{This is not a real restriction as the renormalization group steps of Sec.\ \ref{sec:RG} alternate
between modulated infection and healing rates.}. 
They are arranged
on the bonds of the chain according to the generalized Fibonacci sequences discussed in the
last section. An example ($k=2$) is shown in Fig.\ \ref{fig:chain}.
For $\lambda_A=\lambda_B$, the system coincides with the usual (clean) one-dimensional contact process.

\subsection{Harris-Luck criterion}
\label{subsec:Harris}

Luck \cite{Luck93a} derived a criterion for the stability of a clean critical point
against weak aperiodic inhomogeneities. The basic idea is analogous to that of the
Harris criterion for random disorder: The clean critical point is stable of the
fluctuations $\Delta r$ of the local distance from criticality between different correlation
volumes are smaller than the global distance $r$ to criticality. For aperiodic
inhomogeneities characterized by a wandering exponent $\omega$, the fluctuations
behave as $\Delta r \sim N^{\omega-1} \sim \xi^{d(\omega-1)}$ while the global
distance from criticality scales as $r \sim \xi^{-1/\nu_\perp}$. The condition $\Delta r < r$
for $\xi \to \infty$ leads to the exponent inequality
\begin{equation}
\omega < 1 - \frac 1 {d\nu_\perp}~.
\label{eq:Harris-Luck}
\end{equation}
This is the Harris-Luck criterion. In the case of random disorder, $\omega=1/2$, it reduces
to the usual Harris criterion \cite{Harris74}. If the inequality (\ref{eq:Harris-Luck}) is
fulfilled, weak inhomogeneities are irrelevant, otherwise they are relevant and change
the character of the phase transition.

The correlation length exponent of the one-dimensional clean contact process takes the
value $\nu_\perp \approx 1.097$. The Harris-Luck criterion thus simplifies to
$\omega < 1-1/\nu_\perp \approx 0.0884$. This implies that aperiodic fluctuations of the
transition rates are irrelevant
for $k=1$ and 2 while they are relevant for $k\ge 3$.

\section{Real-space renormalization group}
\label{sec:RG}
\subsection{Overview}
\label{subsec:RG_overview}

This section is devoted to a real-space renormalization group for the contact process on
generalized Fibonacci chains. Our method is inspired by a similar calculation for the
transverse-field Ising chain \cite{FilhoFariaVieira12}. There are, however, some
important differences.

Let us start by assuming that the transition rates fulfill the condition
$\lambda_A \ll \mu \ll \lambda_B$. We can then perform  a renormalization group step
which consists of two parts:

(i) Combine the $(k+1)$ consecutive sites connected by the large $\lambda_B$ infection rate into a
    single new site with a renormalized healing rate $\tilde \mu \ll \mu$. Structurally,
    this reverses one inflation step, as the result is a system with uniform infection rates
    $\lambda_A$ but two different healing rates, $\mu$ and $\tilde \mu$,  modulated
    according to a Fibonacci chain of one generation earlier.

(ii) Integrate out the sites with the original healing rate $\mu$ which is now the largest
    transition rate in the system. This generates renormalized infection rates
    (bonds) $\tilde \lambda$  between
    the remaining sites and reverses another inflation step. The system now has
    uniform healing rate $\tilde \mu$ and two different infection rates $\lambda_A$ and
    $\tilde \lambda$ modulated according to a Fibonacci chain of two generations earlier
    than the original chain.

After renaming $\lambda_A \to \lambda_B, \tilde\lambda \to \lambda_A$ and $\tilde \mu \to \mu$
we arrive at a system equivalent to the original one, but with renormalized transition rates.
As long as the renormalized rates still fulfill the condition $\lambda_A \ll \mu \ll \lambda_B$,
this renormalization group step can be iterated.

In the opposite limit, $\lambda_A \gg \mu \gg \lambda_B$, an analogous renormalization group step
does \emph{not} preserve the structure of the system and can thus not be iterated. However, we will
study the fate of systems in this regime numerically at the end of Sec.\ \ref{sec:MC}. 
If the healing rate $\mu$ is much larger (or smaller) than \emph{both} infection rates, the system
can never reach criticality, instead it is deep in the inactive (or active) phase.

\subsection{Recursion relations}
\label{subsec:RG_recursions}

We now analyze the renormalization group step outlined above in a quantitative manner.
The infection rate $\lambda_B$ is the largest transition rate in the system. Thus, sites
coupled by $\lambda_B$-bonds will quickly reinfect each other when one of them heals.
Consequently, all $k+1$ sites coupled by the $k$ consecutive $\lambda_B$ bonds can be
merged into a single new site of ``moment'' (number of sites)
\begin{equation}
\tilde m = (k+1) m
\label{eq:moment_recursion}
\end{equation}
where $m$ is the moment of the original sites (in the bare system, $m=1$). The renormalized
healing rate $\tilde \mu$ of these new sites can be found either by directly enumerating all
possible healing paths of the cluster or by analyzing the eigenvalues of the generator of the
Markov process in the Hamiltonian formalism (see, e.g., Ref.\ \cite{HooyberghsIgloiVanderzande03,*HooyberghsIgloiVanderzande04}).
Both methods give the same result,
\begin{equation}
\tilde \mu = \alpha_k \frac{\mu^{k+1}}{(\lambda_B/2)^k}
\label{eq:mu_recursion}
\end{equation}
with $\alpha_1=2$, $\alpha_2=4$, and $\alpha_3=8$. If $\mu \ll \lambda_B$, the renormalized healing rate
is strongly reduced, $\tilde\mu \ll \mu$.

After the first part of the renormalization group step, the system has uniform
infection rates $\lambda_A$ but two types of sites, original sites having healing rate $\mu$
and new sites having healing rate $\tilde \mu$. If the rates fulfill the condition
$\tilde \mu \ll \lambda_A \ll \mu$, we can perform the second part of the renormalization group
step and integrate out the original sites which occur in groups of $k$. This leads
to new effective bonds of length $k+1$ and renormalized infection rate
\begin{equation}
\tilde \lambda/2 = \frac {(\lambda_A/2)^{k+1}}{\mu^k}~.
\label{eq:lambda_recursion}
\end{equation}
The renormalization group step is finished after renaming  $\lambda_A \to \lambda_B, \tilde\lambda \to \lambda_A$ and $\tilde \mu \to \mu$.
Equations  (\ref{eq:mu_recursion}) and  (\ref{eq:lambda_recursion}) are similar to the corresponding relations
for the transverse fields and interactions in the transverse-field Ising model on generalized Fibonacci chains
\cite{FilhoFariaVieira12}. The main difference is the extra factor $\alpha_k$ in (\ref{eq:mu_recursion}).

If we now iterate the renormalization group step, we obtain the following recursion relations
\begin{eqnarray}
\lambda_{A,j+1}/2 &=& \frac {(\lambda_{A,j}/2)^{k+1}}{\mu_j^k}~, \qquad \lambda_{B,j+1} = \lambda_{A,j}~,
\label{eq:lambda_j+1}\\
\mu_{j+1} &=& \alpha_k \frac{\mu_j^{k+1}}{(\lambda_{B,j}/2)^k}~,
\label{eq:mu_j+1}\\
m_{j+1} &=& (k+1)\, m_j~,
\label{eq:m_j+1}
\end{eqnarray}
where $j$ is the index of the renormalization group step. For the further analysis, it is convenient to
introduce variables $R_j$ and $S_j$ that characterize the ratios of the transition rates,
\begin{equation}
R_j = \ln(2\mu_j/\lambda_{B,j})~, \qquad S_j = \ln(\lambda_{A,j}/(2\mu_j))~.
\label{eq:RS}
\end{equation}
In terms of these variables, the recursion relations (\ref{eq:lambda_j+1}) and  (\ref{eq:mu_j+1})
turn into an inhomogeneous linear recurrence
\begin{eqnarray}
R_{j+1} &=& \phantom{-} k R_j - S_j + A_k ~,
\label{eq:R_recurrence}\\
S_{j+1} &=& -k R_j +(k+1) S_j -A_k
\label{eq:S_recurrence}
\end{eqnarray}
where $A_k = \ln (\alpha_k)$.

\subsection{Renormalization-group flow}
\label{subsec:RG_flow}

The general solution of the inhomogeneous recurrence (\ref{eq:R_recurrence},\ref{eq:S_recurrence}) is the
sum of a particular solution and the general solution of the corresponding homogeneous recurrence.
To find a particular solution, we use the ansatz $R_j=\bar R =\textrm{const}$ and $S_j=\bar S = \textrm{const}$.
Inserting this into eqs.\ (\ref{eq:R_recurrence}) and (\ref{eq:S_recurrence}) yields
\begin{equation}
\bar R = -\frac {k-1}{k(k-2)}A_k~, \qquad \bar S = -\frac {1}{k(k-2)}A_k~.
\label{eq:RbarSbar}
\end{equation}
The ansatz fails for the case $k=2$ which thus requires a separate calculation. It
will be given in the appendix.

The general solution of the homogeneous recurrence
\begin{eqnarray}
R_{j+1} &=& \phantom{-} k R_j - S_j  ~,
\label{eq:R_homo}\\
S_{j+1} &=& -k R_j +(k+1) S_j
\label{eq:S_homo}
\end{eqnarray}
can be easily found by diagonalizing the coefficient matrix
\begin{equation}
\mathbf{T}_k =  \left(\begin{array}{cc}
                        k & -1 \\
                        -k & k+1 \\
                      \end{array}
                    \right)~.
\label{eq:RGmatrix}
\end{equation}
Its eigenvalues, $\zeta_+^2$ and $\zeta_-^2$, are the squares of the eigenvalues
of the substitution matrix (\ref{eq:subsmatrix}), and the corresponding right eigenvectors
read
\begin{equation}
\left(
       \begin{array}{c}
         1 \\
         -\zeta_+ \\
       \end{array}
     \right)~, \qquad
     \left(
       \begin{array}{c}
         1 \\
         -\zeta_- \\
       \end{array}
     \right)~.
\end{equation}
By decomposing the initial conditions $R_0-\bar R$ and $S_0-\bar S$ into the eigenvectors and
multiplying with the $j$-th power of the matrix $\mathbf{T}_k$, we obtain the solution
\begin{eqnarray}
R_j &=& \frac 1 {\zeta_+ -\zeta_-} \left( -\eta_- \zeta_+^{2j} + \eta_+ \zeta_-^{2j} \right) -\frac {k-1}{k(k-2)}A_k~,~
\label{eq:R_solution}\\
S_j &=& \frac 1 {\zeta_+ -\zeta_-} \left(  \eta_- \zeta_+^{2j+1} - \eta_+ \zeta_-^{2j+1} \right) -\frac {1}{k(k-2)}A_k~.~
\label{eq:S_solution}
\end{eqnarray}
The coefficients $\eta_+$ and $\eta_-$ are determined by the initial ratios $R_0$ and $S_0$ via
\begin{eqnarray}
\eta_\pm = \zeta_\pm R_0 + S_0 + \frac {A_k} {k(k-2)}[1+\zeta_\pm (k-1)]~.
\label{eq:eta}
\end{eqnarray}

Let us analyze the solution (\ref{eq:R_solution}), (\ref{eq:S_solution}) to find the critical point. In the limit $j\to \infty$,
the behavior of $R_j$ and $S_j$ is dominated by the larger of the two eigenvalues as $R_j \sim -\eta_- \zeta_+^{2j}$ and
$S_j \sim \eta_- \zeta_+^{2j+1}$. If $\eta_-$ is negative, $R_j$ flows to $+\infty$ while $S_j$ flows to $-\infty$.
The healing rate $\mu$ thus becomes larger than both infection rates, putting the system into the inactive phase. In contrast,
if $\eta_-$ is positive, $R_j$ flows to $-\infty$ while $S_j$ flows to $+\infty$. In this case, the system is in the active phase
because the healing rate becomes smaller than both infection rates. The critical point is therefore given by the condition
$\eta_-=0$. This can be rewritten in terms of the initial (bare) values of the transition rates as
\begin{equation}
\left( \frac {2\mu}{\lambda_B} \right)^{1-\zeta_-}= \frac {\lambda_A}{\lambda_B} \,\alpha_k^{\frac{1+\zeta_-(k-1)}{k(k-2)}} ~.
\label{eq:phaseboundary}
\end{equation}

\subsection{Critical behavior}
\label{subsec:RG_critical}

At criticality, $\eta_-=0$, the asymptotic behavior of $R_j$ and $S_j$ is determined by the
smaller eigenvalue $\zeta_-$. Specifically,
\begin{eqnarray}
R_j &=& \frac 1 {\zeta_+ -\zeta_-}  \eta_+ \zeta_-^{2j} -\frac {k-1}{k(k-2)}A_k~,~
\label{eq:R_critical}\\
S_j &=& -\frac 1 {\zeta_+ -\zeta_-}   \eta_+ \zeta_-^{2j+1}  -\frac {1}{k(k-2)}A_k~.~
\label{eq:S_critical}
\end{eqnarray}
Both quantities are negative because $\eta_+$ and $\zeta_-$ are negative.
If $|\zeta_-|>1$, both $R_j$ and $S_j$ diverge towards $-\infty$ with increasing $j$, i.e,
the modulation of the transition rates becomes infinitely strong. At the resulting ``infinite-modulation''
critical point, the condition $\lambda_{A,j} \ll \mu_j \ll \lambda_{B,j}$ is better and better fulfilled
with increasing $j$ implying that the renormalization group becomes asymptotically exact.

To determine the critical behavior, we first analyze the flow of the inverse time scale
$\Omega$ under the renormalization group. $\Omega$ can be identified with the largest
transition rate in the system, $\Omega_j = \lambda_{B,j}$. Its recursion relation thus
reads
\begin{equation}
\frac {\Omega_j}{\Omega_{j-1}}=\frac{\lambda_{A,j-1}}{\lambda_{B,j-1}}= \exp(R_{j-1} + S_{j-1})~.
\label{eq:Omega_recursion}
\end{equation}
Inserting the critical solutions (\ref{eq:R_critical}) and (\ref{eq:S_critical}), and iterating
the recursion gives
\begin{equation}
\Omega_j = \alpha_k^{j/(k-2)} \exp\left[\frac{\eta_+(1-\zeta_-)(1-\zeta_-^{2j})}{(\zeta_+ - \zeta_-)(1-\zeta_-^2)}\right] \Omega_0
\label{eq:Omega_solution}
\end{equation}

To relate the inverse time scale $\Omega_j$ to the length scale $\ell_j$, we recall that the length of the generalized
Fibonacci chain increases as $N \sim \zeta_+^i$ with inflation step $i$. As each renormalization group step
corresponds to two inflation steps, this means that the length scale $\ell_j$ behaves as $\ell_j \sim N \sim \zeta_+^{2j}$.
Inserting this relation into (\ref{eq:Omega_solution}), we obtain activated dynamical scaling
of the form
\begin{equation}
\ln(\Omega_0/\Omega_j) \sim \ell_j^\psi~.
\label{eq:activated}
\end{equation}
The tunneling exponent is identical to the wandering exponent of the underlying Fibonacci
chain, i.e., it takes the value
\begin{equation}
\psi = \omega = {\ln|\zeta_-|} / {\ln \zeta_+}~.
\label{eq:tunneling}
\end{equation}

We now turn to the decay of the density $\rho$ of active sites with time at criticality.
Sites (clusters) that survive the renormalization group to step $j$, survive the real time evolution
to time $t_j \sim 1/\Omega_j$. The density of sites after renormalization group step $j$ is easily estimated
as $\rho_j = n_j m_j$ where $n_j \sim 1/\ell_j$ is the density of surviving clusters and $m_j =(k+1)^j$
is their moment. Combining this with eq.\ (\ref{eq:Omega_solution}),
we obtain
\begin{equation}
\rho(t_j) \sim \left[\ln(t_j/t_0)\right]^{-\bar\delta}
\label{eq:rho_t}
\end{equation}
with the critical exponent given by
\begin{equation}
\bar\delta = \frac 1 \psi -\phi = \frac 1 \psi - \frac {\ln(k+1)}{2\ln|\zeta_-|}~.
\label{eq:delta}
\end{equation}
($\phi$ characterizes the relation between cluster moment and inverse time scale,
$m_j \sim [\ln(\Omega_0/\Omega_j)]^\phi$.)

Experiments starting from a single active seed site embedded in an otherwise inactive system
can be characterized by the survival probability $P_s$ and the average number $N_s$ of sites
in the active cloud. Within the renormalization group approach, a run survives to time $t$
if the seed site belongs to a cluster surviving at renormalization scale $\Omega \sim 1/t$. As the
density of (original) sites surviving after renormalization group step $j$ is given by $n_j m_j$,
we find that the survival probability decays with the same critical exponent as the density,
$P_s(t_j) \sim \left[\ln(t_j/t_0)\right]^{-\bar\delta}$. In each of the surviving runs, the number
of infected sites is simply the current size of the renormalization group cluster. Thus,
$N_s(t_j) = n_j m_j^2$. Expressing $j$ in terms of the time scale yields
\begin{equation}
N_s(t_j) \sim \left[\ln(t_j/t_0)\right]^{\bar\Theta}
\label{eq:Ns_t}
\end{equation}
with the so-called critical initial slip exponent given by
\begin{equation}
\bar\Theta = -\frac 1 \psi +2\phi = -\frac 1 \psi + \frac {\ln(k+1)}{\ln|\zeta_-|}~.
\label{eq:Theta}
\end{equation}
Note that $\bar\Theta$, $\bar\delta$ and $\psi$ fulfill the hyperscaling relation
$\bar\Theta+2\bar\delta-1/\psi=0$.

Finally, we turn to the off-critical behavior. Consider a system slightly on the inactive side of
the transition, $\eta_-<0$. According to the general solution (\ref{eq:R_solution}), $R_j$ increases
under renormalization. The character of the flow changes from critical to
that of the inactive phase when $R_j$ reaches 0. This happens at the crossover step $j^*$.
If both $\zeta_+>1$ and $|\zeta_-|>1$, the constant term in (\ref{eq:R_solution}) can be neglected.
This yields a crossover step
\begin{equation}
j^* = \frac 1 2 ~\frac {\ln(\eta_-/\eta_+)}{\ln|\zeta_-/\zeta_+|}~.
\label{eq:j_star}
\end{equation}
The corresponding crossover length scale is given by $\ell_{j^*} \sim \zeta_+^{2j^*} \sim \eta_-^{-\nu_\perp}$
with the correlation length critical exponent
\begin{equation}
\nu_\perp = \frac 1 {1-\psi} =\frac {\ln(\zeta_+)}{\ln(\zeta_+) - \ln|\zeta_-|}~.
\label{eq:nu}
\end{equation}
Interestingly, $\nu_\perp$ exactly saturates the Harris-Luck inequality (\ref{eq:Harris-Luck}).

The critical exponents $\psi, \bar\delta$ and $\nu$ (or, alternatively, $\psi, \bar\Theta$ and $\nu$)
constitute a complete set of exponents. All other exponents can therefore be calculated
from scaling relations, for example, $\beta = \bar\delta \nu_\perp \psi$.

\subsection{Log-periodic oscillations}
\label{subsec:oscillations}

If the renormalized transition rates of consecutive renormalization group steps are well
separated, $\mu_{j+1} \ll \mu_j$ and $\lambda_{A,j+1} \ll \lambda_{A,j}$, the time evolution
of the system proceeds in pronounced steps. For example, each downward step in density of active sites
is associated  with a time given by one of the renormalized decays rates,
$1/t \sim \mu_j$.

The generalized Fibonacci sequences are invariant under the inflation rules (\ref{eq:inflation}),
i.e., they feature \emph{discrete} scale invariance. The steps in various observables are
manifestations of the log-periodic oscillations usually associated with
such discrete scale invariance (see, e.g., Ref.\ \cite{Sornette98} for a review).

Within the real-space renormalization group approach, the steps can be analyzed by comparing
the values of an observable at two consecutive renormalization group steps. The density $\rho$
of active sites and the survival probability $P_s$ behave as $\rho_j \sim P_{s,j} \sim m_j/\ell_j \sim (k+1)^j \zeta_+^{-2j}$.
The step in $\ln \rho$ and $\ln P_s$ is therefore given by
\begin{equation}
\Delta \ln(\rho) = \Delta \ln(P_s) = \ln[(k+1)/\zeta_+^2]~.
\label{eq:rho_step}
\end{equation}
Because of the activated scaling, the oscillations are not log-periodic but double-log periodic
in time, i.e.,
\begin{equation}
\Delta \ln[\ln(t/t_0)] = 2 \ln |\zeta_-|~.
\label{eq:time_step}
\end{equation}
The size $N_s$ of the active cluster growing out of a single seed has analogous steps of
magnitude
\begin{equation}
\Delta \ln(N_s)  = \ln[(k+1)^2/\zeta_+^2]~.
\label{eq:Ns_step}
\end{equation}

\subsection{Explicit predictions for $k=1, 2$ and $3$}
\label{subsec:k=123}

We now apply the general renormalization group theory developed above to the specific
cases $k=1, 2$ and 3.

\emph{$k=1$: Fibonacci chain.} The eigenvalues of the substitution matrix $\mathbf{M}_1$
are given by $\zeta_{\pm} = (1\pm \sqrt{5})/2$. Their numerical values are $\zeta_+ = 1.618$ and
$\zeta_-= -0.6180$. As $|\zeta_-|<1$, the (logarithmic) ratio variables $R_j$ and $S_j$ at criticality do
not approach $-\infty$ under the renormalization group. Instead $R_j$ approaches 0 and $S_j$
goes to a constant. The renormalized transition rates thus eventually violate the condition
$\lambda_A \ll \mu \ll \lambda_B$ (even if the bare rates fulfill it). This implies that the
renormalization group method does not describe the correct asymptotic critical behavior
for $k=1$.

\emph{$k=2$:} The eigenvalues of the substitution matrix $\mathbf{M}_2$ are
$\zeta_+=2$ and $\zeta_-=-1$. As $|\zeta_-|=1$, the system is right at the boundary
between the renormalization group method working and failing, and a more
detailed analysis is required. Although the general solution
(\ref{eq:R_solution},\ref{eq:S_solution}) is not valid for $k=2$,
we have solved this case in the appendix. At criticality, both $R_j$ and $S_j$
go towards large positive values with $j\to \infty$. This means that the renormalization
group method eventually fails for $k=2$ even if the bare inhomogeneities are strong.

\emph{$k=3$:} The substitution matrix $\mathbf{M}_3$ has eigenvalues $\zeta_\pm = (1 \pm \sqrt{13})/2$
with numerical values $\zeta_+ = 2.303$ and $\zeta_- = -1.303$. Because $|\zeta_-|>1$, the
renormalization group is asymptotically exact in this case. Inserting $\zeta_+$ and $\zeta_-$
into eqs.\ (\ref{eq:tunneling}), (\ref{eq:delta}), (\ref{eq:Theta}), and (\ref{eq:nu}), we
obtain the following values for the critical exponents: $\psi=\omega_3=0.3171$, $\bar\delta=0.5330$,
$\bar\Theta=2.086$, and $\nu=1.464$. The steps in the observables can be determined from
eqs.\ (\ref{eq:rho_step}) to (\ref{eq:Ns_step}) yielding $\Delta \ln(\rho)=\Delta \ln(P_s)=0.2819$,
$\Delta \ln(N_s) = 1.104$, and $\Delta \ln[\ln(t/t_0)]=0.5290$.

\emph{$k\ge 4$:} Because $|\zeta_-|$ increases with increasing $k$, the renormalization group method
is valid for all $k\ge 4$. Critical exponents and step sizes can be calculated analogously to the
$k=3$ case.

\section{Monte-Carlo simulations}
\label{sec:MC}
\subsection{Simulation method and overview}
\label{subsec:MC_overview}

To test the predictions of the Harris-Luck criterion and the renormalization group
approach of Sec.\ \ref{sec:RG}, we performed extensive Monte-Carlo simulations.
 Our system is characterized by three transition rates, the
uniform healing rate $\mu$ and the infection rates $\lambda_A$ and $\lambda_B$
which are modulated according to the generalized Fibonacci chain. We set the healing rate
to $\mu=1$ and tune the transition by changing $\lambda_B$.
The ratio $\lambda_A/\lambda_B$ is treated as a fixed external parameter that determines the strength
of the aperiodic inhomogeneity.

Our numerical implementation of the contact process is similar to Ref.\ \cite{Dickman99} but adapted
to the case of nonuniform infection rates. The algorithm starts at time $t=0$ from
some configuration of infected  and healthy sites and consists of a sequence of events. During each event
an infected site is randomly chosen from a list of all $N_a$ infected sites, then a process is selected,
either healing with probability $1/[1+ \max(\lambda_A,\lambda_B)]$, infection of the left
neighbor with probability $(1/2)\lambda_\textrm{left}/[1+ \max(\lambda_A,\lambda_B)]$ or infection of the right
neighbor with probability $(1/2)\lambda_\textrm{right}/[1+ \max(\lambda_A,\lambda_B)]$.
($\lambda_\textrm{left}$ and $\lambda_\textrm{right}$ denote the infection rates of the bonds left and right
of the given site.) The infection succeeds if this neighbor
is healthy. The time is then incremented by $1/N_a$.

Employing this algorithm, we studied the cases $k=1, 2$ and 3 using systems of up to
35 generations of the generalized Fibonacci chain (more than $10^7$ sites). We used several different values of the
parameter characterizing the strength of the inhomogeneity, $\lambda_A/\lambda_B = 0.001,~ 0.004,~ 0.01,~ 0.04,~ 0.1,~ 2/3,~ 1$,
and 25. To cope with the slow dynamics at criticality, we simulated long times up to $1.4 \times 10^9$. Most of our simulations were
spreading runs that start from a single infected seed site and measure the survival probability $P_s$
and the size $N_s$ of the active cloud. The data are averaged over up to 500,000 trials.
For comparison, we have also performed a few density decay runs that start from a fully active lattice.

\subsection{Results for $k=1$}
\label{subsec:MC_k=1}

According to the Harris-Luck criterion, weak inhomogeneities are irrelevant in the $k=1$ case because
the wandering exponent $\omega_1=-1$ fulfills the inequality
$\omega < 1-1/\nu_\perp \approx 0.0884$. Moreover, the renormalization group  of Sec.\
\ref{sec:RG} predicts that strong inhomogeneities decrease under renormalization.
We therefore expect the contact process to display clean DP critical behavior even for strong bare inhomogeneities.

\begin{figure}
\includegraphics[width=8.2cm]{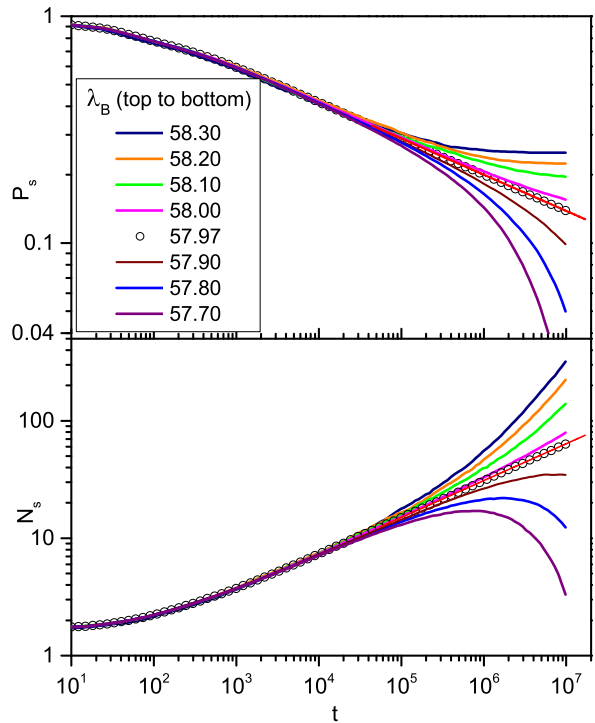}
\caption{(Color online) Survival probability $P_s$ and size $N_s$ of the active cloud vs.\ time $t$
for $k=1$ and strong  inhomogeneity $\lambda_A/\lambda_B=0.01$. The data are averages
of 100,000 (away from criticality) to 500,000 (at criticality) trials. The critical point is located at
$\lambda_B = 57.97$. The solid straight lines represent power-law fits giving the critical exponents
$\delta=0.160$ and $\Theta=0.314$. }
\label{fig:NsPs_vs_t_k=1}
\end{figure}

To test this prediction, we performed spreading simulations of a system
having strong inhomogeneities characterized by $\lambda_A/\lambda_B=0.01$.
The resulting survival probability $P_s$ and size $N_s$ of the active cloud
are presented in Fig.\ \ref{fig:NsPs_vs_t_k=1}.
The figure shows that the critical behavior is of conventional power-law type.
The critical exponents extracted from fits to $P_s \sim t^{-\delta}$ and $N_s \sim t^\Theta$
take the values $\delta=0.160$ and $\Theta=0.314$ in excellent agreement with the clean DP
values $\delta_{\textrm{DP}}=0.159464 $ and $\Theta_{\textrm{DP}}=0.313686$ \cite{Jensen99}.
We thus conclude that the contact process with aperiodic transition rates  modulated according to the $k=1$ Fibonacci
chain is indeed in the clean DP universality class. The same conclusion was reached in Ref.\
\cite{FariaRibeiroSalinas08} based on simulations of the steady-state density $\rho$
for smaller systems.

\subsection{Results for $k=2$}
\label{subsec:MC_k=2}

The wandering exponent $\omega_2=0$ fulfills the Harris-Luck criterion
$\omega < 1-1/\nu_\perp \approx 0.0884$, but just barely. This implies that the inhomogeneities
are asymptotically irrelevant but their magnitude will decrease
only slowly with increasing length scale. The same picture also emerges from the
renormalization group solution given in the appendix: If the bare inhomogeneities
are strong, the renormalization group works for a number of steps until the rates
leave the region of validity $\lambda_A \ll \mu \ll \lambda_B$.
For strong bare inhomogeneities, we therefore expect unconventional behavior
in a transient time regime while the asymptotic behavior should be in the DP
universality class. For sufficiently weak inhomogeneities, the transient regime
will be missing.

\begin{figure}
\includegraphics[width=8.2cm]{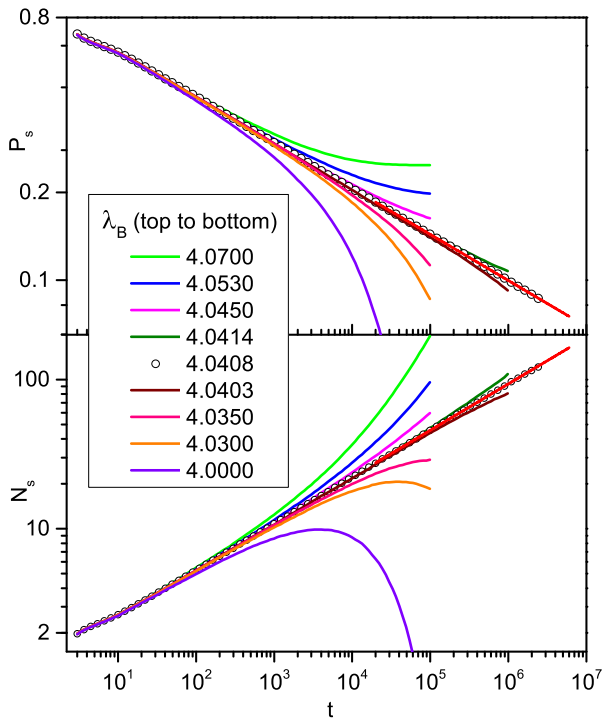}
\caption{(Color online) Survival probability $P_s$ and size $N_s$ of the active cloud
vs. time $t$ for the case $k=2$ and weak inhomogeneity $\lambda_A/\lambda_B=2/3$. The data are averages
of 100,000 to 150,000 trials. The critical point is located at
$\lambda_B = 4.0408$. The solid straight lines represent power-law fits giving the critical exponents
$\delta=0.158$ and $\Theta=0.311$.}
\label{fig:NsPs_vs_t_k=2}
\end{figure}

To verify these predictions, we performed spreading simulations for two different
inhomogeneity strengths, $\lambda_A/\lambda_B=0.01$ and 2/3.
Figure \ref{fig:NsPs_vs_t_k=2} shows the survival probability $P_s$ and size $N_s$ of the active cloud
for the weak inhomogeneity case.
The figure yields conventional power-law critical behavior with exponents
$\delta=0.158$ and $\Theta=0.311$ in excellent agreement with the clean DP
values $\delta_{\textrm{DP}}=0.159464 $ and $\Theta_{\textrm{DP}}=0.313686$.

\begin{figure}
\includegraphics[width=8.2cm]{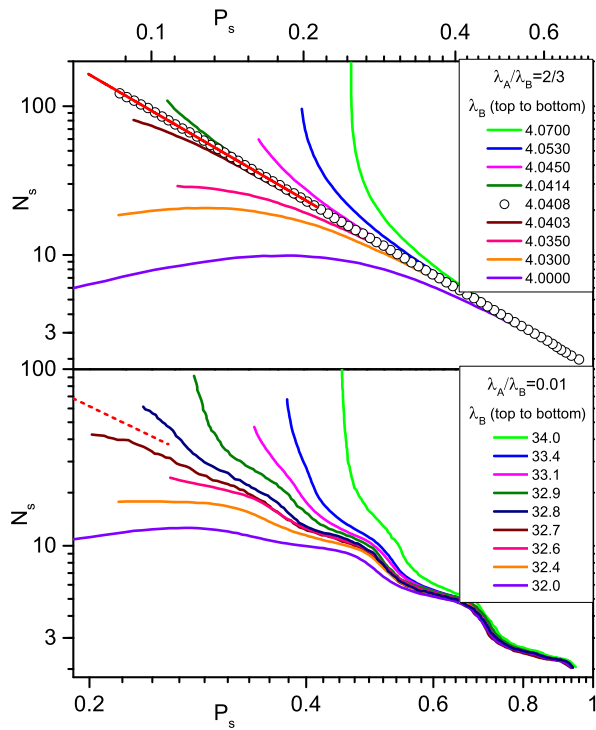}
\caption{(Color online) $N_s$ versus $P_s$ for the case $k=2$ and two different
inhomogeneity strengths,  $\lambda_A/\lambda_B=2/3$ (upper panel) and
$\lambda_A/\lambda_B=0.01$ (lower panel). The data are averages
of 100,000 to 150,000 trials. The solid line in the upper panel is a power-law fit
of the critical curve
($\lambda_B=4.0408$) yielding $\Theta/\delta=1.971$. The dashed line in the lower panel represents
a power law with the clean exponent $-\Theta_{\textrm{DP}}/\delta_{\textrm{DP}}=-1.96712$. }
\label{fig:Ns_vs_Ps_k=2}
\end{figure}

In the case of strong inhomogeneities,  $\lambda_A/\lambda_B=0.01$, the behavior at early times
is different as both $N_s$ and $P_s$ feature oscillations reminiscent of the steps discussed
in Sec.\ \ref{subsec:oscillations}. This becomes particularly clear if one plots $N_s$ vs $P_s$ as
is done in Fig.\ \ref{fig:Ns_vs_Ps_k=2}.
The strength of the oscillations decreases with time, but only slowly.
Therefore, we have not been able to
reach the asymptotic regime within the available simulation times. However, the emerging
critical behavior  for  $\lambda_A/\lambda_B=0.01$ is compatible with the clean DP universality class, as indicated
by the dashed line in the lower panel of
Fig.\ \ref{fig:Ns_vs_Ps_k=2}.

To summarize, we conclude that the asymptotic critical behavior of the $k=2$ chain is
in clean DP universality class for weak inhomogeneities. The same is likely true for strong
inhomogeneities. However, the asymptotic behavior
is approached very slowly, giving rise to an extended transient regime of unconventional behavior
that is controlled by the real-space renormalization group.

\begin{figure*}
\includegraphics[width=18cm]{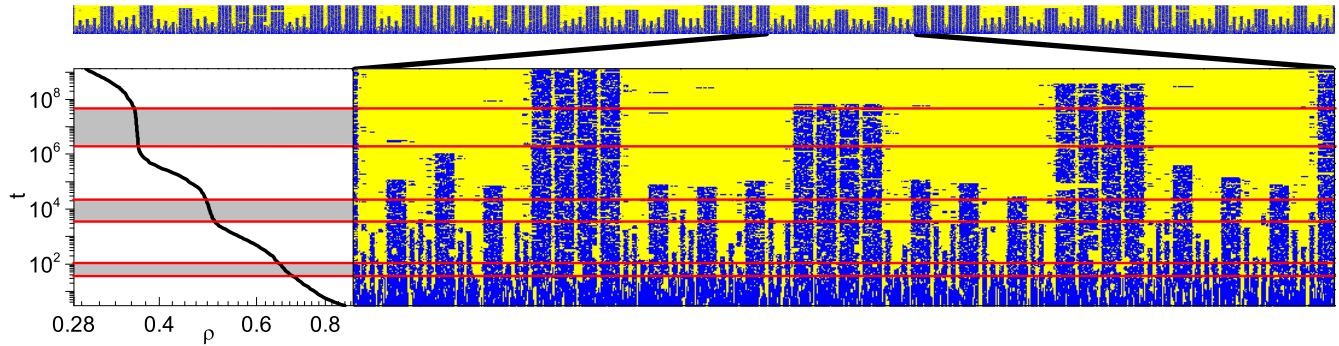}
\caption{(Color online) Example of a density decay run, starting from a fully active
lattice of 15 generations of the $k=3$ chain (173383 sites). The inhomogeneity strength
is $\lambda_A/\lambda_B=0.04$. In the main panel, dark blue dots denote active sites while
light yellow marks inactive sites. The left panel shows the corresponding density $\rho$
of active sites. The horizontal lines are located at times that correspond to the inverse
transition rates at different renormalization group steps, $t=\lambda_B^{-1}, ~ \mu^{-1}$. }
\label{fig:evolution}
\end{figure*}

\subsection{Results for $k=3$}
\label{subsec:MC_k=3}

We now turn to the case of $k=3$ for which the aperiodic inhomogeneities are relevant
according to the Harris-Luck criterion. Moreover, the renormalization group theory
predicts activated dynamical scaling and log-periodic or double-log periodic oscillations in various
observables.

Figure \ref{fig:evolution} shows an example of a density decay run starting from
a fully active lattice for an inhomogeneity strength of $\lambda_A/\lambda_B=0.04$.
The figure clearly illustrates the structure of the time evolution as
the system forms a hierarchy of clusters of active sites that are modulated according to the underlying
$k=3$ generalized Fibonacci sequence. The corresponding time evolution of the density $\rho$ of
active sites progresses in steps; in contrast to the $k=2$ case the steps become sharper and
more pronounced with increasing time $t$.

\begin{figure}
\includegraphics[width=8.2cm]{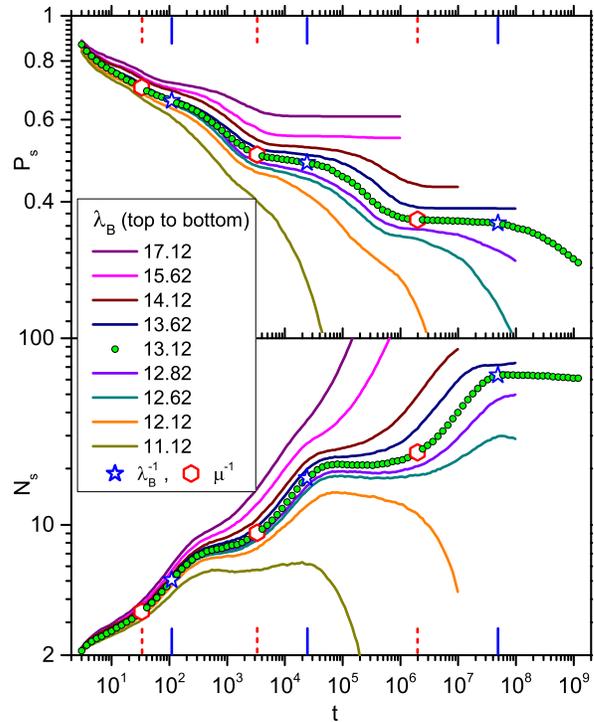}
\caption{(Color online) Survival probability $P_s$ and size $N_s$ of the active cloud
vs.\ time $t$ for the case $k=3$ and
$\lambda_A/\lambda_B=0.04$ (5000 trials).
The steps and plateaus in the critical curve, $\lambda_B=13.12$, become more
pronounced with increasing time. They can be associated with the discrete values
of $\lambda$ and $\mu$ appearing in the renormalization group (marked by large stars and hexagons).}
\label{fig:NsPs_vs_t_k=3}
\end{figure}

\begin{figure}
\includegraphics[width=8.2cm]{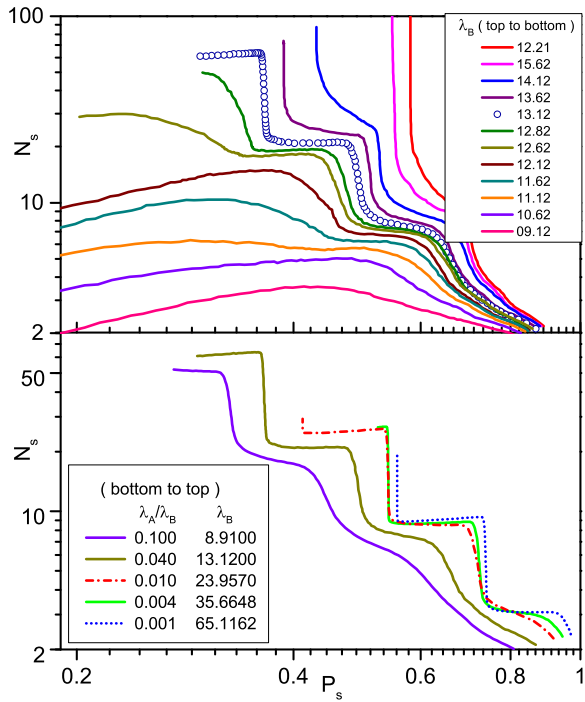}
\caption{(Color online) Upper panel: $N_s$ versus $P_s$ for the case $k=3$ and
$\lambda_A/\lambda_B=0.04$ (5000 trials).
The maximum time is $t_{\textrm{max}}=1.4\times 10^9$ at criticality.
The critical curve, $\lambda_B=13.12$, shows
pronounced steps as predicted in Sec.\ \ref{subsec:oscillations}.
Lower panel: Critical curves for several inhomogeneity strengths
$\lambda_A/\lambda_B$ (5000 to 100000 trials).}
\label{fig:Ns_vs_Ps_k=3}
\end{figure}

To analyze the case $k=3$ quantitatively, we performed extensive spreading runs.
Figure \ref{fig:NsPs_vs_t_k=3} shows the time evolution of $N_s$ and $P_s$
for an inhomogeneity strength $\lambda_A/\lambda_B =0.04$.
Both observables show well-defined steps and plateaus as predicted in Sec.\ \ref{subsec:oscillations}.
They can also be seen in the upper panel of Fig.\  \ref{fig:Ns_vs_Ps_k=3} which shows $N_s$ vs. $P_s$.
In contrast to the $k=2$ case, the steps become more pronounced with increasing time.
Moreover, they can be directly associated with the discrete values of the healing and
infection rates appearing in the renormalization group.

\begin{figure}
\includegraphics[width=8.2cm]{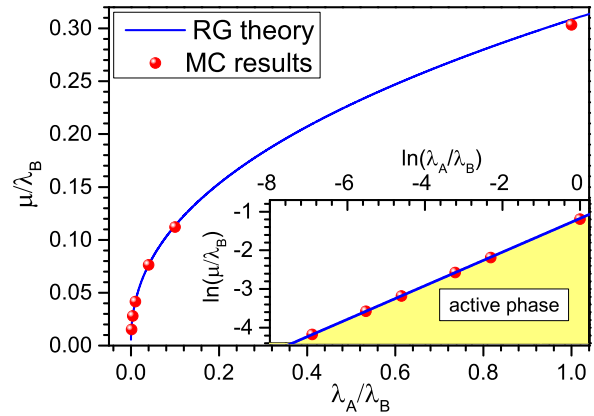}
\caption{(Color online) Phase diagram of the contact process for $k=3$.  The dots are the Monte-Carlo results
  for $\lambda_A/\lambda_B = 0.001$, 0.004, 0.01, 0.04, 0.1, and 1. The solid line represents the renormalization
  group result (\ref{eq:phaseboundary}). }
\label{fig:pd}
\end{figure}

From the upper panel of Fig.\  \ref{fig:Ns_vs_Ps_k=3},
the critical infection rate can be easily found. The critical data feature
well-defined steps and plateaus while the subcritical
and supercritical data curve away from the critical line as predicted.
We performed analogous simulations for inhomogeneity strengths
$\lambda_A/\lambda_B = 0.001$, 0.004, 0.01, 0.1 and 1. The resulting phase diagram
is shown in Fig.\ \ref{fig:pd}.
The Monte-Carlo data are in excellent agreement with the renormalization group prediction
(\ref{eq:phaseboundary}) for all $\lambda_A/\lambda_B \le 0.1$ (even though the analytical result
does not contain any adjustable parameters). Surprisingly, the analytical result
is still a good approximation in the uniform case $\lambda_A/\lambda_B =1$ where the
renormalization group cannot be expected to work.

The effect of the inhomogeneity strength on the critical behavior is demonstrated in the lower panel of Fig.\
\ref{fig:Ns_vs_Ps_k=3} which shows $N_s$ vs.\ $P_s$ for several values of $\lambda_A/\lambda_B$.
If the (bare) inhomogeneities are very strong (small value of $\lambda_A/\lambda_B$),
the steps in the critical $N_s$ vs $P_s$ curve are sharp and pronounced from the outset
because the renormalization group is always in its asymptotic regime $\lambda_A \ll \mu \ll \lambda_B$.
For weaker inhomogeneities, the oscillations of the $N_s$ vs $P_s$ curves are initially
not very pronounced. With increasing time the steps become sharper because the renormalization
group flows towards the asymptotic regime.

\begin{figure}
\includegraphics[width=8.2cm]{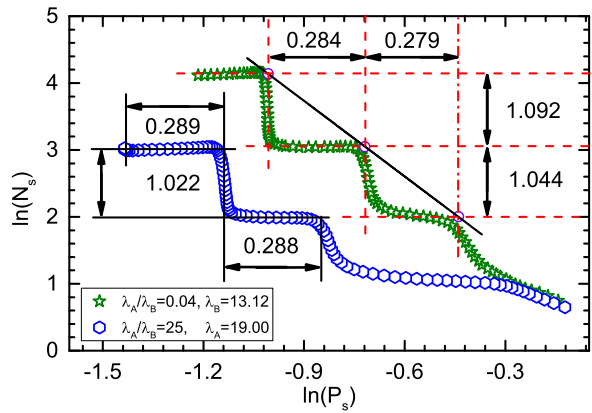}
\caption{(Color online) Quantitative analysis of the critical $N_s$ versus $P_s$ curve for $k=3$ and
$\lambda_A/\lambda_B=0.04$ (maximum time $t_{\textrm{max}}=1.4\times 10^9$) and $\lambda_A/\lambda_B=25$
(maximum time $t_{\textrm{max}}=2\times 10^8$).  The solid line is a fit of the envelop of the curve
to the power law $N_s \sim \ P_s^{-\bar\Theta/\bar\delta}$ yielding $\bar\Theta/\bar\delta =3.79$.}
\label{fig:Ns_vs_Ps_k=3_fit}
\end{figure}

To compare the Monte-Carlo data and the renormalization group predictions quantitatively,
we now investigate the critical $N_s$ vs.\ $P_s$ curve for $\lambda_A/\lambda_B=0.04$ in detail.
The exponent $\bar\Theta/\bar\delta$ can be found by fitting the envelope of the
$N_s$ vs.\ $P_s$ curve. This means fitting equivalent discrete points, each representing
one renormalization group step. This analysis, shown in Fig.\ \ref{fig:Ns_vs_Ps_k=3_fit}
yields $\bar\Theta/\bar\delta =3.79$.
This value is in good agreement with the prediction of 3.91, in particular in view of the fact that
we only have 3 steps to perform the fit.
Figure  \ref{fig:Ns_vs_Ps_k=3_fit} also allows us to determine the steps $\Delta \ln(P_s)$ and
$\Delta \ln (N_s)$ between consecutive renormalization group steps. Using the data of the third step
which is the last complete step in our data, we find $\Delta \ln(P_s)=0.284$ and $\Delta \ln (N_s)=1.092$,
again in good agreement with the renormalization group predictions of Sec.\ \ref{subsec:oscillations},
0.2819 and 1.104, respectively

The renormalization group results of Sec.\ \ref{sec:RG} were derived under the assumption
$\lambda_A \ll \mu \ll \lambda_B$. It is important to investigate whether the resulting
renormalization group fixed point attracts the flow from a larger part of parameter space.
In other words, is the asymptotic critical behavior controlled by this fixed point even if
the bare system violates the condition $\lambda_A \ll \mu \ll \lambda_B$.  In the lower panel of Fig.\
\ref{fig:Ns_vs_Ps_k=3}, we have seen that the fixed point attracts the flow from
regions where $\lambda_A/\lambda_B$ is only moderately small. We now look at an extreme case
in which the bare system strongly violates the condition.  The second curve in Fig.\
 \ref{fig:Ns_vs_Ps_k=3_fit}  shows the critical $N_s$ vs $P_s$
data for $\lambda_A/\lambda_B =25$.
As expected, the two curves initially behave differently. However, the steps $\Delta \ln(P_s)$ and$\Delta \ln(N_s)$
forming at later times appear to be identical within the numerical errors. Moreover, we also performed density decay
runs for $\lambda_A/\lambda_B =25$. A figure (not shown) analogous to Fig.\ \ref{fig:evolution} clearly demonstrates
that the same hierarchy of clusters forms at late times for $\lambda_A/\lambda_B =0.04$ and $\lambda_A/\lambda_B =25$.
This strongly suggests that the
renormalization group fixed point discovered in Sec.\ \ref{sec:RG} also describes the critical behavior
of the system with $\lambda_A/\lambda_B =25$.

\section{Conclusions}
\label{sec:Conclusions}

In summary, we have studied the one-dimensional contact process with aperiodically modulated
transition rates by means of a real-space renormalization group and by Monte-Carlo simulations.
We have focused on modulations according to three cases $k=1,2,3$ of the generalized Fibonacci
sequence defined by the inflation rules A $\to$ AB$^k$ and B $\to$ A. For $k=1$ (the Fibonacci
chain proper), the inhomogeneities are strongly irrelevant according to the Harris-Luck
criterion at the clean DP critical point. Correspondingly, our numerical simulations yield
critical behavior in the clean DP universality class even if the initial inhomogeneities
are strong. This agrees with earlier results on the steady state density \cite{FariaRibeiroSalinas08}.

In the $k=2$ case, the inhomogeneities are still irrelevant at the clean DP critical point,
but just barely. This implies that their scale dimension is close to zero. The inhomogeneity strength
therefore decreases only slowly with increasing length and time scales. Our Monte-Carlo simulations
confirm this picture. If the (bare) inhomogeneities are weak, we again find critical behavior in the
clean DP universality class. For strong inhomogeneities, the system shows unconventional behavior
in an extended transient time regime that is controlled by the real-space renormalization group.
The long-time evolution appears to approach the clean DP critical behavior. However, we could not
reach the true asymptotic regime within our simulation time for strong inhomogeneities.

For $k\ge 3$, the aperiodic modulation of the transition rates is relevant at the clean DP critical
point because the Harris-Luck criterion is violated.
We have developed a renormalization group theory of the transition and identified a fixed point
that describes unconventional criticality. At this infinite-modulation fixed point, the inhomogeneity strength diverges,
and the method becomes asymptotically exact.
The resulting critical behavior is characterized by activated dynamical scaling.
Moreover, the time dependence of observables such as the density of active sites, the survival probability, and the size
of the active cloud show striking plateaus and steps. They are a consequence of the discrete scale
invariance of the generalized Fibonacci sequence and related to the log-periodic oscillations
found in many aperiodic systems (see, e.g., \cite{Sornette98}). Due to the activated dynamical scaling, the oscillations
are actually \emph{double-log} periodic in time. Analogous double-log oscillations should occur in
other systems featuring activated scaling, for example in quantum spin chains \cite{Vieira05}.

We have numerically confirmed these renormalization group predictions
for the case $k=3$. The Monte-Carlo simulations also provide evidence for the critical behavior to be universal,
i.e., it is valid for both weak and strong aperiodic modulations.

It is interesting to compare the phase transitions in the aperiodic contact process (for $k\ge 3$)
and the disordered contact process. In both cases, the fluctuations of the transition rates at criticality
diverge with increasing length scale. In the disordered contact process, this leads to an infinite-randomness
critical point \cite{HooyberghsIgloiVanderzande03,*HooyberghsIgloiVanderzande04,IgloiMonthus05}; and for
aperiodic rates, the critical point is of infinite-modulation type.  Both these
critical points feature unconventional activated dynamical scaling rather than the usual power-law scaling.
In the disordered case, the critical point is accompanied by Griffiths singularities
\cite{Griffiths69,Noest86,*Noest88,Vojta06} which are missing in the aperiodic case because the
generalized Fibonacci chains do not have rare regions. Conversely, the log-periodic oscillations
of observables in the aperiodic chain do not exist in the disordered chain because the latter
system has continuous rather than discrete scale invariance (in the statistical sense).

Our renormalization group method is similar to the approach used in Ref.\ \cite{FilhoFariaVieira12}
to study the aperiodic transverse-field Ising chain. In fact, the critical behavior of the
contact process and the transverse-field Ising chain are identical in the cases in which the
renormalization group correctly describes the critical point (i.e., $k\ge 3$). This mirrors the behavior
of the corresponding random systems: The random transverse-field Ising chain \cite{Fisher95} and
the random contact process \cite{HooyberghsIgloiVanderzande03,*HooyberghsIgloiVanderzande04}
feature the same critical exponents.

The main difference between the Ising chain and the contact process
occurs for $k=2$. For the Ising chain, $k=2$ aperiodic modulations are exactly marginal according to
the Harris-Luck criterion. This is reflected in the fact that the modulation strength stays constant
under the renormalization group, leading to nonuniversal critical behavior \cite{FilhoFariaVieira12}.
In contrast, $k=2$ aperiodic modulations of the contact process are weakly irrelevant. Correspondingly,
the renormalization group works at best in a transient time regime while the asymptotic critical behavior
appears to be in the clean DP universality class.

Recently, aperiodic sequences were used to construct complex networks with long-range connections; and the
contact process on such networks was studied \cite{JuhaszOdor09}. The nonequilibrium  phase transition
features power-law critical behavior with exponents that depend on the underlying network.
Time-dependent quantities exhibit log-periodic oscillations due to the discrete scale invariance
of the networks.

Let us finally comment on generalizations to higher dimensions. One could, for example, construct higher-dimensional
aperiodic modulations of the transition rates by repeating identical one-dimensional sequences in
the second (and third) direction. This would increase the relevance of the modulations in the Harris-Luck criterion
because the clean correlation length exponent decreases with increasing dimension while the
fluctuations of the distance to criticality are unchanged. In the random case, such correlated inhomogeneities
lead to a smearing of the DP critical point \cite{Vojta04,*DickisonVojta05} because rare regions undergo the transition
independently. As the aperiodic systems do not have any rare regions, their behavior is likely different
Alternatively, one could also look at more general tilings
in two and three dimensions. Of particular interest are structures with unbounded fluctuations
such as the tiling proposed in Ref.\ \cite{GodrecheLancon92}. Studying the contact process on such
lattices remains a task for future.

\section*{Acknowledgements}

This work has been supported in part by the NSF under grants no. DMR-0906566,
DMR-1205803, and PHYS-1066293.

\section*{Appendix: The case $k=2$}
\label{sec:appendix}

The general solution of the renormalization group developed in Sec.\ \ref{subsec:RG_flow}
does not apply to the case $k=2$ because the particular solution of the inhomogeneous
recurrence (\ref{eq:R_recurrence}) and (\ref{eq:S_recurrence}) is not of the form
$R_j=\bar R =\textrm{const}$ and $S_j=\bar S = \textrm{const}$. The reason is that
the smaller eigenvalue of the coefficient matrix $\mathbf{T}_2$ takes the value $\zeta_-^2=1$.

In this appendix, we therefore directly solve the problem for $k=2$. After introducing
the variables $X_j=2 R_j + S_j$ and $Y_j=R_j -S_j$ into (\ref{eq:R_recurrence}) and (\ref{eq:S_recurrence}),
the recurrence relations read
\begin{eqnarray}
X_{j+1} &=& X_j + A ~,\label{eq:X_recurrence}\\
Y_{j+1} &=& 4 Y_j +2 A~. \label{eq:Y_recurrence}
\end{eqnarray}
As the two equations are now decoupled, they can be easily solved,
\begin{eqnarray}
X_j &=& X_0 + j A \label{eq:X_solution} \\
Y_j &=& -\frac 2 3 A + 4^j \left(Y_0 +\frac 2 3 A \right) \label{eq:Y_solution}
\end{eqnarray}
where $X_0 =2R_0 + S_0$ and $Y_0=R_0 -S_0$.
Transforming back to the variables $R_j$ and $S_j$, we finally obtain
\begin{eqnarray}
R_j &=& \frac 1 3 \left[ X_0 + j A - \frac 2 3 A + 4^j \left( Y_0 + \frac 2 3 A \right) \right]~,
\label{eq:R_solution_k2} \\
S_j &=& \frac 1 3 \left[ X_0 + j A + \frac 4 3 A - 2\times 4^j \left( Y_0 + \frac 2 3 A \right) \right]~.
\label{eq:S_solution_k2}
\end{eqnarray}
For $Y_0 + 2A/3>0$, the system is in the inactive phase because $R_j \to \infty$ and $S_j \to -\infty$
under the renormalization group. In contrast, the system is in the active phase for $Y_0 + 2A/3<0$.
At criticality, $Y_0 + 2A/3 = 0$, both $R_j$ and $S_j$ increase linearly with $j$. The implies that the
renormalization group method asymptotically fails because the transition rates eventually violate the condition
$\lambda_A \ll \mu \ll \lambda_B$.

\bibliographystyle{apsrev4-1}
\bibliography{../00Bibtex/rareregions}

\begin{thebibliography}{50}%
\makeatletter
\providecommand \@ifxundefined [1]{%
 \@ifx{#1\undefined}
}%
\providecommand \@ifnum [1]{%
 \ifnum #1\expandafter \@firstoftwo
 \else \expandafter \@secondoftwo
 \fi
}%
\providecommand \@ifx [1]{%
 \ifx #1\expandafter \@firstoftwo
 \else \expandafter \@secondoftwo
 \fi
}%
\providecommand \natexlab [1]{#1}%
\providecommand \enquote  [1]{``#1''}%
\providecommand \bibnamefont  [1]{#1}%
\providecommand \bibfnamefont [1]{#1}%
\providecommand \citenamefont [1]{#1}%
\providecommand \href@noop [0]{\@secondoftwo}%
\providecommand \href [0]{\begingroup \@sanitize@url \@href}%
\providecommand \@href[1]{\@@startlink{#1}\@@href}%
\providecommand \@@href[1]{\endgroup#1\@@endlink}%
\providecommand \@sanitize@url [0]{\catcode `\\12\catcode `\$12\catcode
  `\&12\catcode `\#12\catcode `\^12\catcode `\_12\catcode `\%12\relax}%
\providecommand \@@startlink[1]{}%
\providecommand \@@endlink[0]{}%
\providecommand \url  [0]{\begingroup\@sanitize@url \@url }%
\providecommand \@url [1]{\endgroup\@href {#1}{\urlprefix }}%
\providecommand \urlprefix  [0]{URL }%
\providecommand \Eprint [0]{\href }%
\providecommand \doibase [0]{http://dx.doi.org/}%
\providecommand \selectlanguage [0]{\@gobble}%
\providecommand \bibinfo  [0]{\@secondoftwo}%
\providecommand \bibfield  [0]{\@secondoftwo}%
\providecommand \translation [1]{[#1]}%
\providecommand \BibitemOpen [0]{}%
\providecommand \bibitemStop [0]{}%
\providecommand \bibitemNoStop [0]{.\EOS\space}%
\providecommand \EOS [0]{\spacefactor3000\relax}%
\providecommand \BibitemShut  [1]{\csname bibitem#1\endcsname}%
\let\auto@bib@innerbib\@empty
\bibitem [{\citenamefont {Zhdanov}\ and\ \citenamefont
  {Kasemo}(1994)}]{ZhdanovKasemo94}%
  \BibitemOpen
  \bibfield  {author} {\bibinfo {author} {\bibfnamefont {V.~P.}\ \bibnamefont
  {Zhdanov}}\ and\ \bibinfo {author} {\bibfnamefont {B.}~\bibnamefont
  {Kasemo}},\ }\href@noop {} {\bibfield  {journal} {\bibinfo  {journal}
  {Surface Science Reports}\ }\textbf {\bibinfo {volume} {20}},\ \bibinfo
  {pages} {113} (\bibinfo {year} {1994})}\BibitemShut {NoStop}%
\bibitem [{\citenamefont {Schmittmann}\ and\ \citenamefont
  {Zia}(1995)}]{SchmittmannZia95}%
  \BibitemOpen
  \bibfield  {author} {\bibinfo {author} {\bibfnamefont {B.}~\bibnamefont
  {Schmittmann}}\ and\ \bibinfo {author} {\bibfnamefont {R.~K.~P.}\
  \bibnamefont {Zia}},\ }in\ \href@noop {} {\emph {\bibinfo {booktitle} {Phase
  Transitions and Critical Phenomena}}},\ Vol.~\bibinfo {volume} {17},\
  \bibinfo {editor} {edited by\ \bibinfo {editor} {\bibfnamefont
  {C.}~\bibnamefont {Domb}}\ and\ \bibinfo {editor} {\bibfnamefont {J.~L.}\
  \bibnamefont {Lebowitz}}}\ (\bibinfo  {publisher} {Academic},\ \bibinfo
  {address} {New York},\ \bibinfo {year} {1995})\ p.~\bibinfo {pages}
  {1}\BibitemShut {NoStop}%
\bibitem [{\citenamefont {Marro}\ and\ \citenamefont
  {Dickman}(1999)}]{MarroDickman99}%
  \BibitemOpen
  \bibfield  {author} {\bibinfo {author} {\bibfnamefont {J.}~\bibnamefont
  {Marro}}\ and\ \bibinfo {author} {\bibfnamefont {R.}~\bibnamefont
  {Dickman}},\ }\href@noop {} {\emph {\bibinfo {title} {Nonequilibrium Phase
  Transitions in Lattice Models}}}\ (\bibinfo  {publisher} {Cambridge
  University Press},\ \bibinfo {address} {Cambridge},\ \bibinfo {year}
  {1999})\BibitemShut {NoStop}%
\bibitem [{\citenamefont {Hinrichsen}(2000)}]{Hinrichsen00}%
  \BibitemOpen
  \bibfield  {author} {\bibinfo {author} {\bibfnamefont {H.}~\bibnamefont
  {Hinrichsen}},\ }\href {\doibase 10.1080/00018730050198152} {\bibfield
  {journal} {\bibinfo  {journal} {Adv. Phys.}\ }\textbf {\bibinfo {volume}
  {49}},\ \bibinfo {pages} {815} (\bibinfo {year} {2000})}\BibitemShut
  {NoStop}%
\bibitem [{\citenamefont {Odor}(2004)}]{Odor04}%
  \BibitemOpen
  \bibfield  {author} {\bibinfo {author} {\bibfnamefont {G.}~\bibnamefont
  {Odor}},\ }\href {\doibase 10.1103/RevModPhys.76.663} {\bibfield  {journal}
  {\bibinfo  {journal} {Rev. Mod. Phys.}\ }\textbf {\bibinfo {volume} {76}},\
  \bibinfo {pages} {663} (\bibinfo {year} {2004})}\BibitemShut {NoStop}%
\bibitem [{\citenamefont {L{\"u}beck}(2004)}]{Luebeck04}%
  \BibitemOpen
  \bibfield  {author} {\bibinfo {author} {\bibfnamefont {S.}~\bibnamefont
  {L{\"u}beck}},\ }\href@noop {} {\bibfield  {journal} {\bibinfo  {journal}
  {Int. J. Mod. Phys. B}\ }\textbf {\bibinfo {volume} {18}},\ \bibinfo {pages}
  {3977} (\bibinfo {year} {2004})}\BibitemShut {NoStop}%
\bibitem [{\citenamefont {T{\"a}uber}\ \emph {et~al.}(2005)\citenamefont
  {T{\"a}uber}, \citenamefont {Howard},\ and\ \citenamefont
  {Vollmayr-Lee}}]{TauberHowardVollmayrLee05}%
  \BibitemOpen
  \bibfield  {author} {\bibinfo {author} {\bibfnamefont {U.~C.}\ \bibnamefont
  {T{\"a}uber}}, \bibinfo {author} {\bibfnamefont {M.}~\bibnamefont {Howard}},
  \ and\ \bibinfo {author} {\bibfnamefont {B.~P.}\ \bibnamefont
  {Vollmayr-Lee}},\ }\href@noop {} {\bibfield  {journal} {\bibinfo  {journal}
  {J. Phys. A}\ }\textbf {\bibinfo {volume} {38}},\ \bibinfo {pages} {R79}
  (\bibinfo {year} {2005})}\BibitemShut {NoStop}%
\bibitem [{\citenamefont {Henkel}\ \emph {et~al.}(2008)\citenamefont {Henkel},
  \citenamefont {Hinrichsen},\ and\ \citenamefont
  {L{\"u}beck}}]{HenkelHinrichsenLuebeck_book08}%
  \BibitemOpen
  \bibfield  {author} {\bibinfo {author} {\bibfnamefont {M.}~\bibnamefont
  {Henkel}}, \bibinfo {author} {\bibfnamefont {H.}~\bibnamefont {Hinrichsen}},
  \ and\ \bibinfo {author} {\bibfnamefont {S.}~\bibnamefont {L{\"u}beck}},\
  }\href@noop {} {\emph {\bibinfo {title} {Non-equilibrium phase transitions.
  Vol 1: Absorbing phase transitions}}}\ (\bibinfo  {publisher} {Springer},\
  \bibinfo {address} {Dordrecht},\ \bibinfo {year} {2008})\BibitemShut
  {NoStop}%
\bibitem [{\citenamefont {Grassberger}\ and\ \citenamefont {de~la
  Torre}(1979)}]{GrassbergerdelaTorre79}%
  \BibitemOpen
  \bibfield  {author} {\bibinfo {author} {\bibfnamefont {P.}~\bibnamefont
  {Grassberger}}\ and\ \bibinfo {author} {\bibfnamefont {A.}~\bibnamefont
  {de~la Torre}},\ }\href {\doibase 10.1016/0003-4916(79)90207-0} {\bibfield
  {journal} {\bibinfo  {journal} {Ann. Phys. (NY)}\ }\textbf {\bibinfo {volume}
  {122}},\ \bibinfo {pages} {373} (\bibinfo {year} {1979})}\BibitemShut
  {NoStop}%
\bibitem [{\citenamefont {Janssen}(1981)}]{Janssen81}%
  \BibitemOpen
  \bibfield  {author} {\bibinfo {author} {\bibfnamefont {H.~K.}\ \bibnamefont
  {Janssen}},\ }\href@noop {} {\bibfield  {journal} {\bibinfo  {journal} {Z.
  Phys. B}\ }\textbf {\bibinfo {volume} {42}},\ \bibinfo {pages} {151}
  (\bibinfo {year} {1981})}\BibitemShut {NoStop}%
\bibitem [{\citenamefont {Grassberger}(1982)}]{Grassberger82}%
  \BibitemOpen
  \bibfield  {author} {\bibinfo {author} {\bibfnamefont {P.}~\bibnamefont
  {Grassberger}},\ }\href@noop {} {\bibfield  {journal} {\bibinfo  {journal}
  {Z. Phys. B}\ }\textbf {\bibinfo {volume} {47}},\ \bibinfo {pages} {365}
  (\bibinfo {year} {1982})}\BibitemShut {NoStop}%
\bibitem [{\citenamefont {Harris}(1974{\natexlab{a}})}]{HarrisTE74}%
  \BibitemOpen
  \bibfield  {author} {\bibinfo {author} {\bibfnamefont {T.~E.}\ \bibnamefont
  {Harris}},\ }\href {\doibase doi:10.1214/aop/1176996493} {\bibfield
  {journal} {\bibinfo  {journal} {Ann. Prob.}\ }\textbf {\bibinfo {volume}
  {2}},\ \bibinfo {pages} {969} (\bibinfo {year}
  {1974}{\natexlab{a}})}\BibitemShut {NoStop}%
\bibitem [{\citenamefont {Takeuchi}\ \emph {et~al.}(2007)\citenamefont
  {Takeuchi}, \citenamefont {Kuroda}, \citenamefont {Chate},\ and\
  \citenamefont {Sano}}]{TKCS07}%
  \BibitemOpen
  \bibfield  {author} {\bibinfo {author} {\bibfnamefont {K.~A.}\ \bibnamefont
  {Takeuchi}}, \bibinfo {author} {\bibfnamefont {M.}~\bibnamefont {Kuroda}},
  \bibinfo {author} {\bibfnamefont {H.}~\bibnamefont {Chate}}, \ and\ \bibinfo
  {author} {\bibfnamefont {M.}~\bibnamefont {Sano}},\ }\href@noop {} {\bibfield
   {journal} {\bibinfo  {journal} {Phys. Rev. Lett.}\ }\textbf {\bibinfo
  {volume} {99}},\ \bibinfo {pages} {234503} (\bibinfo {year}
  {2007})}\BibitemShut {NoStop}%
\bibitem [{\citenamefont {Corte}\ \emph {et~al.}(2008)\citenamefont {Corte},
  \citenamefont {Chaikin}, \citenamefont {Gollub},\ and\ \citenamefont
  {Pine}}]{CCGP08}%
  \BibitemOpen
  \bibfield  {author} {\bibinfo {author} {\bibfnamefont {L.}~\bibnamefont
  {Corte}}, \bibinfo {author} {\bibfnamefont {P.~M.}\ \bibnamefont {Chaikin}},
  \bibinfo {author} {\bibfnamefont {J.~P.}\ \bibnamefont {Gollub}}, \ and\
  \bibinfo {author} {\bibfnamefont {D.~J.}\ \bibnamefont {Pine}},\ }\href@noop
  {} {\bibfield  {journal} {\bibinfo  {journal} {Nature Physics}\ }\textbf
  {\bibinfo {volume} {4}},\ \bibinfo {pages} {420} (\bibinfo {year}
  {2008})}\BibitemShut {NoStop}%
\bibitem [{\citenamefont {Franceschini}\ \emph {et~al.}(2011)\citenamefont
  {Franceschini}, \citenamefont {Filippidi}, \citenamefont {Guazzelli},\ and\
  \citenamefont {Pine}}]{FFGP11}%
  \BibitemOpen
  \bibfield  {author} {\bibinfo {author} {\bibfnamefont {A.}~\bibnamefont
  {Franceschini}}, \bibinfo {author} {\bibfnamefont {E.}~\bibnamefont
  {Filippidi}}, \bibinfo {author} {\bibfnamefont {E.}~\bibnamefont
  {Guazzelli}}, \ and\ \bibinfo {author} {\bibfnamefont {D.~J.}\ \bibnamefont
  {Pine}},\ }\href {\doibase 10.1103/PhysRevLett.107.250603} {\bibfield
  {journal} {\bibinfo  {journal} {Phys. Rev. Lett.}\ }\textbf {\bibinfo
  {volume} {107}},\ \bibinfo {pages} {250603} (\bibinfo {year}
  {2011})}\BibitemShut {NoStop}%
\bibitem [{\citenamefont {Okuma}\ \emph {et~al.}(2011)\citenamefont {Okuma},
  \citenamefont {Tsugawa},\ and\ \citenamefont
  {Motohashi}}]{OkumaTsugawaMotohashi11}%
  \BibitemOpen
  \bibfield  {author} {\bibinfo {author} {\bibfnamefont {S.}~\bibnamefont
  {Okuma}}, \bibinfo {author} {\bibfnamefont {Y.}~\bibnamefont {Tsugawa}}, \
  and\ \bibinfo {author} {\bibfnamefont {A.}~\bibnamefont {Motohashi}},\ }\href
  {\doibase 10.1103/PhysRevB.83.012503} {\bibfield  {journal} {\bibinfo
  {journal} {Phys. Rev. B}\ }\textbf {\bibinfo {volume} {83}},\ \bibinfo
  {pages} {012503} (\bibinfo {year} {2011})}\BibitemShut {NoStop}%
\bibitem [{\citenamefont {Janssen}(1997)}]{Janssen97}%
  \BibitemOpen
  \bibfield  {author} {\bibinfo {author} {\bibfnamefont {H.~K.}\ \bibnamefont
  {Janssen}},\ }\href@noop {} {\bibfield  {journal} {\bibinfo  {journal} {Phys.
  Rev. E}\ }\textbf {\bibinfo {volume} {55}},\ \bibinfo {pages} {6253}
  (\bibinfo {year} {1997})}\BibitemShut {NoStop}%
\bibitem [{\citenamefont {Harris}(1974{\natexlab{b}})}]{Harris74}%
  \BibitemOpen
  \bibfield  {author} {\bibinfo {author} {\bibfnamefont {A.~B.}\ \bibnamefont
  {Harris}},\ }\href {\doibase 10.1088/0022-3719/7/9/009} {\bibfield  {journal}
  {\bibinfo  {journal} {J. Phys. C}\ }\textbf {\bibinfo {volume} {7}},\
  \bibinfo {pages} {1671} (\bibinfo {year} {1974}{\natexlab{b}})}\BibitemShut
  {NoStop}%
\bibitem [{\citenamefont {Bramson}\ \emph {et~al.}(1991)\citenamefont
  {Bramson}, \citenamefont {Durrett},\ and\ \citenamefont
  {Schonmann}}]{BramsonDurrettSchonmann91}%
  \BibitemOpen
  \bibfield  {author} {\bibinfo {author} {\bibfnamefont {M.}~\bibnamefont
  {Bramson}}, \bibinfo {author} {\bibfnamefont {R.}~\bibnamefont {Durrett}}, \
  and\ \bibinfo {author} {\bibfnamefont {R.~H.}\ \bibnamefont {Schonmann}},\
  }\href@noop {} {\bibfield  {journal} {\bibinfo  {journal} {Ann. Prob.}\
  }\textbf {\bibinfo {volume} {19}},\ \bibinfo {pages} {960} (\bibinfo {year}
  {1991})}\BibitemShut {NoStop}%
\bibitem [{\citenamefont {Moreira}\ and\ \citenamefont
  {Dickman}(1996)}]{MoreiraDickman96}%
  \BibitemOpen
  \bibfield  {author} {\bibinfo {author} {\bibfnamefont {A.~G.}\ \bibnamefont
  {Moreira}}\ and\ \bibinfo {author} {\bibfnamefont {R.}~\bibnamefont
  {Dickman}},\ }\href@noop {} {\bibfield  {journal} {\bibinfo  {journal} {Phys.
  Rev. E}\ }\textbf {\bibinfo {volume} {54}},\ \bibinfo {pages} {R3090}
  (\bibinfo {year} {1996})}\BibitemShut {NoStop}%
\bibitem [{\citenamefont {Dickman}\ and\ \citenamefont
  {Moreira}(1998)}]{DickmanMoreira98}%
  \BibitemOpen
  \bibfield  {author} {\bibinfo {author} {\bibfnamefont {R.}~\bibnamefont
  {Dickman}}\ and\ \bibinfo {author} {\bibfnamefont {A.~G.}\ \bibnamefont
  {Moreira}},\ }\href@noop {} {\bibfield  {journal} {\bibinfo  {journal} {Phys.
  Rev. E}\ }\textbf {\bibinfo {volume} {57}},\ \bibinfo {pages} {1263}
  (\bibinfo {year} {1998})}\BibitemShut {NoStop}%
\bibitem [{\citenamefont {Webman}\ \emph {et~al.}(1998)\citenamefont {Webman},
  \citenamefont {Avraham}, \citenamefont {Cohen},\ and\ \citenamefont
  {Havlin}}]{WACH98}%
  \BibitemOpen
  \bibfield  {author} {\bibinfo {author} {\bibfnamefont {I.}~\bibnamefont
  {Webman}}, \bibinfo {author} {\bibfnamefont {D.~B.}\ \bibnamefont {Avraham}},
  \bibinfo {author} {\bibfnamefont {A.}~\bibnamefont {Cohen}}, \ and\ \bibinfo
  {author} {\bibfnamefont {S.}~\bibnamefont {Havlin}},\ }\href@noop {}
  {\bibfield  {journal} {\bibinfo  {journal} {Phil. Mag. B}\ }\textbf {\bibinfo
  {volume} {77}},\ \bibinfo {pages} {1401} (\bibinfo {year}
  {1998})}\BibitemShut {NoStop}%
\bibitem [{\citenamefont {Cafiero}\ \emph {et~al.}(1998)\citenamefont
  {Cafiero}, \citenamefont {Gabrielli},\ and\ \citenamefont
  {Munoz}}]{CafieroGabrielliMunoz98}%
  \BibitemOpen
  \bibfield  {author} {\bibinfo {author} {\bibfnamefont {R.}~\bibnamefont
  {Cafiero}}, \bibinfo {author} {\bibfnamefont {A.}~\bibnamefont {Gabrielli}},
  \ and\ \bibinfo {author} {\bibfnamefont {M.~A.}\ \bibnamefont {Munoz}},\
  }\href@noop {} {\bibfield  {journal} {\bibinfo  {journal} {Phys. Rev. E}\
  }\textbf {\bibinfo {volume} {57}},\ \bibinfo {pages} {5060} (\bibinfo {year}
  {1998})}\BibitemShut {NoStop}%
\bibitem [{\citenamefont {Hooyberghs}\ \emph {et~al.}(2003)\citenamefont
  {Hooyberghs}, \citenamefont {Igl\'oi},\ and\ \citenamefont
  {Vanderzande}}]{HooyberghsIgloiVanderzande03}%
  \BibitemOpen
  \bibfield  {author} {\bibinfo {author} {\bibfnamefont {J.}~\bibnamefont
  {Hooyberghs}}, \bibinfo {author} {\bibfnamefont {F.}~\bibnamefont {Igl\'oi}},
  \ and\ \bibinfo {author} {\bibfnamefont {C.}~\bibnamefont {Vanderzande}},\
  }\href {\doibase 10.1103/PhysRevLett.90.100601} {\bibfield  {journal}
  {\bibinfo  {journal} {Phys. Rev. Lett.}\ }\textbf {\bibinfo {volume} {90}},\
  \bibinfo {pages} {100601} (\bibinfo {year} {2003})}\BibitemShut {NoStop}%
\bibitem [{\citenamefont {Hooyberghs}\ \emph {et~al.}(2004)\citenamefont
  {Hooyberghs}, \citenamefont {Igl\'oi},\ and\ \citenamefont
  {Vanderzande}}]{HooyberghsIgloiVanderzande04}%
  \BibitemOpen
  \bibfield  {author} {\bibinfo {author} {\bibfnamefont {J.}~\bibnamefont
  {Hooyberghs}}, \bibinfo {author} {\bibfnamefont {F.}~\bibnamefont {Igl\'oi}},
  \ and\ \bibinfo {author} {\bibfnamefont {C.}~\bibnamefont {Vanderzande}},\
  }\href@noop {} {\bibfield  {journal} {\bibinfo  {journal} {Phys. Rev. E}\
  }\textbf {\bibinfo {volume} {69}},\ \bibinfo {pages} {066140} (\bibinfo
  {year} {2004})}\BibitemShut {NoStop}%
\bibitem [{\citenamefont {Noest}(1986)}]{Noest86}%
  \BibitemOpen
  \bibfield  {author} {\bibinfo {author} {\bibfnamefont {A.~J.}\ \bibnamefont
  {Noest}},\ }\href@noop {} {\bibfield  {journal} {\bibinfo  {journal} {Phys.
  Rev. Lett.}\ }\textbf {\bibinfo {volume} {57}},\ \bibinfo {pages} {90}
  (\bibinfo {year} {1986})}\BibitemShut {NoStop}%
\bibitem [{\citenamefont {Noest}(1988)}]{Noest88}%
  \BibitemOpen
  \bibfield  {author} {\bibinfo {author} {\bibfnamefont {A.~J.}\ \bibnamefont
  {Noest}},\ }\href@noop {} {\bibfield  {journal} {\bibinfo  {journal} {Phys.
  Rev. B}\ }\textbf {\bibinfo {volume} {38}},\ \bibinfo {pages} {2715}
  (\bibinfo {year} {1988})}\BibitemShut {NoStop}%
\bibitem [{\citenamefont {Vojta}\ and\ \citenamefont
  {Dickison}(2005)}]{VojtaDickison05}%
  \BibitemOpen
  \bibfield  {author} {\bibinfo {author} {\bibfnamefont {T.}~\bibnamefont
  {Vojta}}\ and\ \bibinfo {author} {\bibfnamefont {M.}~\bibnamefont
  {Dickison}},\ }\href {\doibase 10.1103/PhysRevE.72.036126} {\bibfield
  {journal} {\bibinfo  {journal} {Phys. Rev. E}\ }\textbf {\bibinfo {volume}
  {72}},\ \bibinfo {pages} {036126} (\bibinfo {year} {2005})}\BibitemShut
  {NoStop}%
\bibitem [{\citenamefont {de~Oliveira}\ and\ \citenamefont
  {Ferreira}(2008)}]{OliveiraFerreira08}%
  \BibitemOpen
  \bibfield  {author} {\bibinfo {author} {\bibfnamefont {M.~M.}\ \bibnamefont
  {de~Oliveira}}\ and\ \bibinfo {author} {\bibfnamefont {S.~C.}\ \bibnamefont
  {Ferreira}},\ }\href@noop {} {\bibfield  {journal} {\bibinfo  {journal} {J.
  Stat. Mech.}\ }\textbf {\bibinfo {volume} {2008}},\ \bibinfo {pages} {P11001}
  (\bibinfo {year} {2008})}\BibitemShut {NoStop}%
\bibitem [{\citenamefont {Vojta}\ \emph {et~al.}(2009)\citenamefont {Vojta},
  \citenamefont {Farquhar},\ and\ \citenamefont {Mast}}]{VojtaFarquharMast09}%
  \BibitemOpen
  \bibfield  {author} {\bibinfo {author} {\bibfnamefont {T.}~\bibnamefont
  {Vojta}}, \bibinfo {author} {\bibfnamefont {A.}~\bibnamefont {Farquhar}}, \
  and\ \bibinfo {author} {\bibfnamefont {J.}~\bibnamefont {Mast}},\ }\href
  {\doibase 10.1103/PhysRevE.79.011111} {\bibfield  {journal} {\bibinfo
  {journal} {Phys. Rev. E}\ }\textbf {\bibinfo {volume} {79}},\ \bibinfo
  {pages} {011111} (\bibinfo {year} {2009})}\BibitemShut {NoStop}%
\bibitem [{\citenamefont {Vojta}(2012)}]{Vojta12}%
  \BibitemOpen
  \bibfield  {author} {\bibinfo {author} {\bibfnamefont {T.}~\bibnamefont
  {Vojta}},\ }\href {\doibase 10.1103/PhysRevE.86.051137} {\bibfield  {journal}
  {\bibinfo  {journal} {Phys. Rev. E}\ }\textbf {\bibinfo {volume} {86}},\
  \bibinfo {pages} {051137} (\bibinfo {year} {2012})}\BibitemShut {NoStop}%
\bibitem [{\citenamefont {Vojta}\ and\ \citenamefont {Lee}(2006)}]{VojtaLee06}%
  \BibitemOpen
  \bibfield  {author} {\bibinfo {author} {\bibfnamefont {T.}~\bibnamefont
  {Vojta}}\ and\ \bibinfo {author} {\bibfnamefont {M.~Y.}\ \bibnamefont
  {Lee}},\ }\href@noop {} {\bibfield  {journal} {\bibinfo  {journal} {Phys.
  Rev. Lett.}\ }\textbf {\bibinfo {volume} {96}},\ \bibinfo {pages} {035701}
  (\bibinfo {year} {2006})}\BibitemShut {NoStop}%
\bibitem [{\citenamefont {Lee}\ and\ \citenamefont {Vojta}(2009)}]{LeeVojta09}%
  \BibitemOpen
  \bibfield  {author} {\bibinfo {author} {\bibfnamefont {M.~Y.}\ \bibnamefont
  {Lee}}\ and\ \bibinfo {author} {\bibfnamefont {T.}~\bibnamefont {Vojta}},\
  }\href@noop {} {\bibfield  {journal} {\bibinfo  {journal} {Phys. Rev. E}\
  }\textbf {\bibinfo {volume} {79}},\ \bibinfo {pages} {041112} (\bibinfo
  {year} {2009})}\BibitemShut {NoStop}%
\bibitem [{\citenamefont {Luck}(1993)}]{Luck93a}%
  \BibitemOpen
  \bibfield  {author} {\bibinfo {author} {\bibfnamefont {J.~M.}\ \bibnamefont
  {Luck}},\ }\href {http://stacks.iop.org/0295-5075/24/i=5/a=007} {\bibfield
  {journal} {\bibinfo  {journal} {EPL (Europhysics Letters)}\ }\textbf
  {\bibinfo {volume} {24}},\ \bibinfo {pages} {359} (\bibinfo {year}
  {1993})}\BibitemShut {NoStop}%
\bibitem [{\citenamefont {Moody}(1997)}]{Moody97}%
  \BibitemOpen
  \bibinfo {editor} {\bibfnamefont {R.}~\bibnamefont {Moody}},\ ed.,\
  \href@noop {} {\emph {\bibinfo {title} {The Mathematics of Long-range
  Aperiodic Order}}}\ (\bibinfo  {publisher} {Kluwer},\ \bibinfo {address}
  {Dordrecht},\ \bibinfo {year} {1997})\BibitemShut {NoStop}%
\bibitem [{Note1()}]{Note1}%
  \BibitemOpen
  \bibinfo {note} {This is not a real restriction as the renormalization group
  steps of Sec.\ \ref {sec:RG} alternate between modulated infection and
  healing rates.}\BibitemShut {Stop}%
\bibitem [{\citenamefont {Filho}\ \emph {et~al.}(2012)\citenamefont {Filho},
  \citenamefont {Faria},\ and\ \citenamefont {Vieira}}]{FilhoFariaVieira12}%
  \BibitemOpen
  \bibfield  {author} {\bibinfo {author} {\bibfnamefont {F.~J.~O.}\
  \bibnamefont {Filho}}, \bibinfo {author} {\bibfnamefont {M.~S.}\ \bibnamefont
  {Faria}}, \ and\ \bibinfo {author} {\bibfnamefont {A.~P.}\ \bibnamefont
  {Vieira}},\ }\href {http://stacks.iop.org/1742-5468/2012/i=03/a=P03007}
  {\bibfield  {journal} {\bibinfo  {journal} {J. Stat. Mech.}\ }\textbf
  {\bibinfo {volume} {2012}},\ \bibinfo {pages} {P03007} (\bibinfo {year}
  {2012})}\BibitemShut {NoStop}%
\bibitem [{\citenamefont {Sornette}(1998)}]{Sornette98}%
  \BibitemOpen
  \bibfield  {author} {\bibinfo {author} {\bibfnamefont {D.}~\bibnamefont
  {Sornette}},\ }\href {\doibase
  http://dx.doi.org/10.1016/S0370-1573(97)00076-8} {\bibfield  {journal}
  {\bibinfo  {journal} {Physics Reports}\ }\textbf {\bibinfo {volume} {297}},\
  \bibinfo {pages} {239 } (\bibinfo {year} {1998})}\BibitemShut {NoStop}%
\bibitem [{\citenamefont {Dickman}(1999)}]{Dickman99}%
  \BibitemOpen
  \bibfield  {author} {\bibinfo {author} {\bibfnamefont {R.}~\bibnamefont
  {Dickman}},\ }\href@noop {} {\bibfield  {journal} {\bibinfo  {journal} {Phys.
  Rev. E}\ }\textbf {\bibinfo {volume} {60}},\ \bibinfo {pages} {R2441}
  (\bibinfo {year} {1999})}\BibitemShut {NoStop}%
\bibitem [{\citenamefont {Jensen}(1999)}]{Jensen99}%
  \BibitemOpen
  \bibfield  {author} {\bibinfo {author} {\bibfnamefont {I.}~\bibnamefont
  {Jensen}},\ }\href@noop {} {\bibfield  {journal} {\bibinfo  {journal} {J.
  Phys. A}\ }\textbf {\bibinfo {volume} {32}},\ \bibinfo {pages} {5233}
  (\bibinfo {year} {1999})}\BibitemShut {NoStop}%
\bibitem [{\citenamefont {Faria}\ \emph {et~al.}(2008)\citenamefont {Faria},
  \citenamefont {Ribeiro},\ and\ \citenamefont
  {Salinas}}]{FariaRibeiroSalinas08}%
  \BibitemOpen
  \bibfield  {author} {\bibinfo {author} {\bibfnamefont {M.~S.}\ \bibnamefont
  {Faria}}, \bibinfo {author} {\bibfnamefont {D.~J.}\ \bibnamefont {Ribeiro}},
  \ and\ \bibinfo {author} {\bibfnamefont {S.~R.}\ \bibnamefont {Salinas}},\
  }\href {http://stacks.iop.org/1742-5468/2008/i=01/a=P01022} {\bibfield
  {journal} {\bibinfo  {journal} {J. Stat. Mech.}\ }\textbf {\bibinfo {volume}
  {2008}},\ \bibinfo {pages} {P01022} (\bibinfo {year} {2008})}\BibitemShut
  {NoStop}%
\bibitem [{\citenamefont {Vieira}(2005)}]{Vieira05}%
  \BibitemOpen
  \bibfield  {author} {\bibinfo {author} {\bibfnamefont {A.~P.}\ \bibnamefont
  {Vieira}},\ }\href {\doibase 10.1103/PhysRevLett.94.077201} {\bibfield
  {journal} {\bibinfo  {journal} {Phys. Rev. Lett.}\ }\textbf {\bibinfo
  {volume} {94}},\ \bibinfo {pages} {077201} (\bibinfo {year}
  {2005})}\BibitemShut {NoStop}%
\bibitem [{\citenamefont {Igloi}\ and\ \citenamefont
  {Monthus}(2005)}]{IgloiMonthus05}%
  \BibitemOpen
  \bibfield  {author} {\bibinfo {author} {\bibfnamefont {F.}~\bibnamefont
  {Igloi}}\ and\ \bibinfo {author} {\bibfnamefont {C.}~\bibnamefont
  {Monthus}},\ }\href@noop {} {\bibfield  {journal} {\bibinfo  {journal} {Phys.
  Rep.}\ }\textbf {\bibinfo {volume} {412}},\ \bibinfo {pages} {277} (\bibinfo
  {year} {2005})}\BibitemShut {NoStop}%
\bibitem [{\citenamefont {Griffiths}(1969)}]{Griffiths69}%
  \BibitemOpen
  \bibfield  {author} {\bibinfo {author} {\bibfnamefont {R.~B.}\ \bibnamefont
  {Griffiths}},\ }\href {\doibase 10.1103/PhysRevLett.23.17} {\bibfield
  {journal} {\bibinfo  {journal} {Phys. Rev. Lett.}\ }\textbf {\bibinfo
  {volume} {23}},\ \bibinfo {pages} {17} (\bibinfo {year} {1969})}\BibitemShut
  {NoStop}%
\bibitem [{\citenamefont {Vojta}(2006)}]{Vojta06}%
  \BibitemOpen
  \bibfield  {author} {\bibinfo {author} {\bibfnamefont {T.}~\bibnamefont
  {Vojta}},\ }\href {\doibase 10.1088/0305-4470/39/22/R01} {\bibfield
  {journal} {\bibinfo  {journal} {J. Phys. A}\ }\textbf {\bibinfo {volume}
  {39}},\ \bibinfo {pages} {R143} (\bibinfo {year} {2006})}\BibitemShut
  {NoStop}%
\bibitem [{\citenamefont {Fisher}(1995)}]{Fisher95}%
  \BibitemOpen
  \bibfield  {author} {\bibinfo {author} {\bibfnamefont {D.~S.}\ \bibnamefont
  {Fisher}},\ }\href {\doibase 10.1103/PhysRevB.51.6411} {\bibfield  {journal}
  {\bibinfo  {journal} {Phys. Rev. B}\ }\textbf {\bibinfo {volume} {51}},\
  \bibinfo {pages} {6411} (\bibinfo {year} {1995})}\BibitemShut {NoStop}%
\bibitem [{\citenamefont {Juh\'asz}\ and\ \citenamefont
  {\'Odor}(2009)}]{JuhaszOdor09}%
  \BibitemOpen
  \bibfield  {author} {\bibinfo {author} {\bibfnamefont {R.}~\bibnamefont
  {Juh\'asz}}\ and\ \bibinfo {author} {\bibfnamefont {G.}~\bibnamefont
  {\'Odor}},\ }\href {\doibase 10.1103/PhysRevE.80.041123} {\bibfield
  {journal} {\bibinfo  {journal} {Phys. Rev. E}\ }\textbf {\bibinfo {volume}
  {80}},\ \bibinfo {pages} {041123} (\bibinfo {year} {2009})}\BibitemShut
  {NoStop}%
\bibitem [{\citenamefont {Vojta}(2004)}]{Vojta04}%
  \BibitemOpen
  \bibfield  {author} {\bibinfo {author} {\bibfnamefont {T.}~\bibnamefont
  {Vojta}},\ }\href@noop {} {\bibfield  {journal} {\bibinfo  {journal} {Phys.
  Rev. E}\ }\textbf {\bibinfo {volume} {70}},\ \bibinfo {pages} {026108}
  (\bibinfo {year} {2004})}\BibitemShut {NoStop}%
\bibitem [{\citenamefont {Dickison}\ and\ \citenamefont
  {Vojta}(2005)}]{DickisonVojta05}%
  \BibitemOpen
  \bibfield  {author} {\bibinfo {author} {\bibfnamefont {M.}~\bibnamefont
  {Dickison}}\ and\ \bibinfo {author} {\bibfnamefont {T.}~\bibnamefont
  {Vojta}},\ }\href@noop {} {\bibfield  {journal} {\bibinfo  {journal} {J.
  Phys. A}\ }\textbf {\bibinfo {volume} {38}},\ \bibinfo {pages} {1199}
  (\bibinfo {year} {2005})}\BibitemShut {NoStop}%
\bibitem [{\citenamefont {Godr{\`e}che}\ and\ \citenamefont
  {Lan\c{c}on}(1992)}]{GodrecheLancon92}%
  \BibitemOpen
  \bibfield  {author} {\bibinfo {author} {\bibfnamefont {C.}~\bibnamefont
  {Godr{\`e}che}}\ and\ \bibinfo {author} {\bibfnamefont {F.}~\bibnamefont
  {Lan\c{c}on}},\ }\href@noop {} {\bibfield  {journal} {\bibinfo  {journal} {J.
  Phys. I France}\ }\textbf {\bibinfo {volume} {2}},\ \bibinfo {pages} {207}
  (\bibinfo {year} {1992})}\BibitemShut {NoStop}%
\end{thebibliography}%
\end{document}